\documentclass[useAMS,usenatbib]{mn2e}

\usepackage{longtable}
\usepackage{multirow}
\usepackage{graphicx,amssymb,amsfonts,amsmath}
\usepackage{subfigure}
\usepackage{natbib}
\usepackage{lscape}

\DeclareRobustCommand{\ion}[2]{%
\relax\ifmmode
\ifx\testbx\f@series
{\mathbf{#1\,\mathsc{#2}}}\else
{\mathrm{#1\,\mathsc{#2}}}\fi
\else\textup{#1\,{\mdseries\textsc{#2}}}%
\fi}

\newcommand{\hst}{{\it HST~}}
\newcommand{\fuse}{{\it FUSE~}}
\newcommand{\ebv}{{$E(B-V)~$}}

\newcommand{\fnz}{f(N_{\textrm{\ion{H}{i}}}, z)}

\newcommand{\Tl}{T_\lambda}
\newcommand{\flf}{f_{\lambda,\rm f}}

\newcommand{\mbh}{M_{\rm BH}}
\newcommand{\lbol}{L_{\rm bol}}
\newcommand{\lumedd}{L_{\rm Edd}}
\newcommand{\nh}{N_{\textrm{\ion{H}{i}}}}

\newcommand{\ledd}{\lambda_{\rm Edd}}

\newcommand{\mi}{\rm M_i(z=2)}

\newcommand{\mryd}{\rm M_{912}}
\newcommand{\mnorm}{\rm M_{1450}}

\newcommand{\mnhi}{N_{\rm HI}}

\newcommand{\mkms}{{\rm km~s^{-1}}}
\newcommand{\rev}[1]{{ #1}}

\title[The first UV quasar stacked spectrum at $z\simeq2.4$ from WFC3]{The first ultraviolet quasar stacked spectrum at $z\simeq2.4$ from WFC3}
\author[E. Lusso et al.]{E.~Lusso$^{1,2}$, G.~ Worseck$^2$, J.~F.~Hennawi$^2$, J.~X.~Prochaska$^3$, C.~Vignali$^4$, J.~Stern$^2$, \newauthor and J.~M.~O'Meara$^5$\\
$^{1}$INAF--Osservatorio Astrofisico di Arcetri, 50125 Florence, Italy.\\
$^{2}$Max Planck Institut f\"{u}r Astronomie, K\"{o}nigstuhl 17, D-69117, Heidelberg, Germany.\\
$^{3}$Department of Astronomy and Astrophysics, UCO/Lick Observatory, University of California, 1156 High Street, Santa Cruz, CA 95064, USA.\\
$^{4}$Dipartimento di Fisica e Astronomia, Universit\`{a} degli Studi di Bologna, Viale Berti-Pichat 6/2, 40127 Bologna, Italy.\\
$^{5}$Department of Chemistry and Physics, Saint Michael's College. One Winooski Park, Colchester, VT 05439, USA.
}

\begin{document}
\date{Draft, \today}

\pagerange{\pageref{firstpage}--\pageref{lastpage}} \pubyear{2002}

\maketitle

\label{firstpage}

\begin{abstract}
The ionising continuum from active galactic nuclei (AGN) is fundamental for interpreting their broad emission lines and understanding their impact on the surrounding gas. Furthermore, it provides hints on how matter accretes onto supermassive black holes. Using {\it HST}'s Wide Field Camera 3 we have constructed the first stacked ultraviolet (rest-frame wavelengths 600--2500\AA)  spectrum of 53 luminous quasars at $z\simeq2.4$, with a state-of-the-art correction for the intervening Lyman forest and Lyman continuum absorption.
The continuum slope ($f_\nu \propto \nu^{\alpha_\nu}$) of the full sample shows a break at $\sim$912\,\AA\ with spectral index $\alpha_\nu=-0.61\pm0.01$ at $\lambda>912$\,\AA\ and a softening at shorter wavelengths ($\alpha_\nu=-1.70 \pm 0.61$ at $\lambda\leq 912$\,\AA). Our analysis proves that a proper intergalactic medium absorption correction is required to establish the intrinsic continuum emission of quasars. We interpret our average ultraviolet spectrum in the context of photoionisation, accretion disk models, and quasar contribution to the ultraviolet background. We find that observed broad line ratios are consistent with those predicted assuming an ionising slope of $\alpha_\mathrm{ion}$$=$$-2.0$, similar to the observed ionising spectrum in the same wavelength range. The continuum break and softening are consistent with accretion disk plus X--ray corona models when black hole spin is taken into account. Our spectral energy distribution yields a 30\% increase to previous estimates of the specific quasar emissivity, such that quasars may contribute significantly to the total specific Lyman limit emissivity estimated from the Ly$\alpha$ forest at $z< 3.2$.
\end{abstract}


\section{Introduction} 
\label{Introduction}

Considerable effort has been devoted to characterising the shape of the quasar (QSO) ionising continuum in the ultraviolet (UV) over the past years. The spectral energy distribution (SED) of active galactic nuclei (AGN) shows a prominent bump, the so-called ``Big Blue Bump" (BBB), which appears to peak in the UV and decline at higher energies \citep{1989ApJ...347...29S,1994ApJS...95....1E}. However, the {\it intrinsic} position and possible luminosity dependence of the BBB has not been properly estimated due to the lack of UV observations corrected for the intergalactic medium (IGM) absorption by neutral hydrogen along the line of sight \citep{2006ApJS..166..470R,2007AJ....133.1780T,2011ApJS..196....2S,2012ApJ...759....6E}. 

One of the main predictions of accretion disc models is that the thermal disc should peak at bluer wavelengths for smaller black hole masses (i.e., $T_{\rm max}\propto (L/\lumedd)^{1/4} \mbh^{-1/4}$; \citealt{1973A&A....24..337S}).
However, both the presence and the position of the break are highly dependent on the IGM correction considered.

From a theoretical perspective, the AGN ionising continuum is crucial
for interpreting broad and narrow emission-lines observed in AGN
spectra, and their relative ratios (see \citealt{2014MNRAS.445.3011S} and \citealt{2014MNRAS.438..604B} for recent models of these emission line regions).  There
is a general consensus that AGN emission lines are produced by a
photoionising continuum (extending from optical-UV to X-ray) which emerges
from an accretion disc around the black hole and by a hot
($T\sim10^{8-9}$ K) plasma of relativistic electrons that Compton
up-scatter the photons coming from the disc \citep{1991ApJ...380L..51H,1993ApJ...413..507H}.  

Quasars are relevant (maybe even the dominant) sources of ionising photons that determine the ionisation state and the temperature of the $z\sim 3$ IGM \citep[e.g.][]{haardt96,haardt12,meiksin03,faucher09}. While not numerous enough at $z\ga 6$ to have contributed significantly to \ion{H}{i} reionisation \citep[e.g.][]{meiksin05,jiang08,shankar07,willott10,fontanot12,fontanot14}, they are likely the only sources responsible for the reionisation of \ion{He}{ii} at $z\sim 3$ \citep{miralda00,faucher08,furlanetto09,mcquinn09,haardt12,compostella13}. All these studies rely on parameterizations of the quasar SED in the BBB region motivated by existing observations, the uncertainties of which are rarely discussed.

From an observational point of view, composite spectra of AGN were constructed by taking advantage of several surveys (LBQS, \citealt{1991ApJ...373..465F}; FIRST, \citealt{2001ApJ...546..775B}; SDSS, \citealt{vandenberk2001}). 
In all these studies, the rest-frame optical composites indicate that the continuum can be described by a power law of the form $f_\nu \propto \nu^{\alpha_\nu}$, with $-0.5\la\alpha_\nu\la -0.3$ in the wavelength range 1200--4000\,\AA.
The first composite that suggests a softening in the far-ultraviolet (blueward of \ion{Ly}{$\alpha$}) was reported by \citet[Z97 hereafter]{zheng97} who analysed 101 quasars from the {\it Hubble Space Telescope} (\hst) in the redshift range $0.33<z<3.6$, covering the wavelengths between 350 and 3000 \AA. 
This softening was interpreted as Comptonization of the thermal disc emission in a hot
corona above the disc \citep{1987ApJ...321..305C}, due to the similarity between the slope of $-$1.8 found by Z97 at $\lambda<1216$\AA~and the one measured by \citet{1997ApJ...477...93L} for a sample of radio quiet X--ray selected quasars ($\alpha_\nu\sim-1.7$), for which simple accretion disc+X--ray corona models were utilised. The break would thus indicate the peak of the BBB.
The work by Z97 has been extended by \citet[hereafter T02]{2002ApJ...565..773T} with more than 80 quasars from \hst over a similar redshift range. This analysis confirmed the findings of Z97 that the UV spectral continuum can be parametrised by a broken power law with the break in the vicinity of the \ion{Ly}{$\alpha$}, with a continuum slope of $\alpha_\nu=-1.57$ for the radio quiet T02 sample.
These results are at variance with those presented by \citet[S04 hereafter]{2004ApJ...615..135S}, who considered more than 100 AGN at $z<0.1$ observed with the {\it Far Ultraviolet Spectroscopic Explorer} (\fuse), covering the rest-frame wavelength range 630$-$1100\AA.
The spectral slope of the \fuse composite spectrum is $\alpha_\nu=-0.56$, significantly harder than previous estimates in the far-infrared from \hst studies. 
\citet[S12 hereafter]{2012ApJ...752..162S} have measured the AGN ionising continua in 22 AGN at $0.026<z<1.44$ using the {\it Cosmic Origins Spectrograph} (COS) on \hst, covering the rest-frame wavelength range 500$-$2000\AA. 
The COS composite shows a break at $\lambda\sim1000$\AA, in line with the previous estimates by Z97 and T02, but with slightly harder spectral index ($\alpha_\nu=-0.68$ at $\lambda=1200-2000$\AA~and $\alpha_\nu=-1.41$ at $\lambda=500-1000$\AA).
 Recently, Stevans et al. (2014, S14 hereafter) considerably improved the S12 analysis by adding 137 AGN to the original sample for a total of 159 sources, selected from the COS archive at redshifts $0.001< z <1.476$. This new COS composite is fully consistent with the one in S12 and shows similar spectral indexes. We note that previous estimates of the break were performed with very few spectra ($\sim$10 in T02, 3-12 in S12, $\sim$20 in S14) contributing at short wavelengths (e.g. $< 700$\AA).

The differences among various surveys may arise from several factors,
such as the small number of observations covering $\lambda<1216$\AA\ (e.g. less than 20 AGN at $z>2$ in the T02 sample), and the crucial placement of the
``continuum windows" in order to construct the quasar ionising
continua.  
Although the latter may not bias significantly the results, 
a possible uncertainty arises from the selection criteria involved
in defining the various samples. The traditional strategy is to
extract any source available from the HST/FUSE archives and apply
several cuts such as redshift, signal-to-noise, and wavelength
coverage. However, the brightest UV sources were usually targeted for
ultraviolet spectroscopy, primarily for absorption line studies.
Furthermore, any archival research has the tendency to include more
peculiar objects that were selected for special investigations, and
thereby the same objects were re--observed with every generation of UV
spectrographs.  These issues likely bias any UV archival sample to be very 
bright in the ultraviolet with a highly heterogeneous selection function.

An additional source of variance comes from the correction for the
IGM absorption employed by different authors.  
Although the technique adopted by T02 and S04 was similar, i.e. a statistical correction for the
unidentified absorbers by considering an empirical parametrisation
\citep[see][]{1993MNRAS.262..499P,2001ApJ...553..528D}, the overall
correction applied was very different as pointed out by S04. In fact, S04
noted that the correction considered by T02 was less than 1\% over the
whole rest-frame wavelength range, which underestimated the
number of \ion{Ly}{$\alpha$} absorbers by a factor of $\sim50$ at
$z=0.1$.  Another simplification usually adopted in previous studies is
the use of a single correction for quasar samples spanning a large range
of redshift, which, because of strong evolution in the IGM absorption, likely
results in significant errors. 

In this paper, we present the first \emph{average} quasar spectrum in the rest-frame wavelength range 600--2500\,\AA\ corrected for intervening \ion{H}{i} Lyman forest and continuum absorption with state-of the-art IGM transmission functions calibrated to the most recent observations \rev{\citep{2014MNRAS.438..476P}}. The sample consists of 53 $z_{\rm em}\simeq 2.4$ quasars from the \hst survey for Lyman limit absorption systems (LLSs) using the Wide Field Camera 3 (WFC3) presented in \citet[O11 hereafter]{2011ApJS..195...16O}.
The structure of this paper is as follows. In Section~\ref{The data-set} we discuss the sample and the selection criteria. In Section~\ref{Composite construction} we describe the technique to construct the stacked spectrum, whilst the IGM transmission curves adopted to correct the observed average spectrum are presented in Section~\ref{IGM transmission correction}, where we also describe our IGM corrected stack with uncertainties. The results and implications of our analysis are discussed in Section~\ref{Results}, while the conclusions are presented in Section~\ref{Summary and Conclusions}.

We adopt a concordance flat $\Lambda$-cosmology with $H_0=70\, \rm{km \,s^{-1}\, Mpc^{-1}}$, $\Omega_\mathrm{m}=0.3$, and $\Omega_\Lambda=0.7$
\citep{komatsu09}. Unless noted otherwise, we will distinguish between the following wavelength ranges in the UV: (i) the near UV (NUV; 2000--3000\,\AA), (ii) the far UV (FUV; 912--2000\,\AA), and (iii) the extreme UV (EUV; $\lambda<912$\,\AA).

\section{The data set}
\label{The data-set}

The data sample employed in the present analysis comes from a survey performed with \hst using 
the WFC3 instrument. 
Quasar sample, \hst observations, and reduction procedures are described in detail in O11. 
In this section we provide a brief summary of this data-set.

The O11 survey consists of 53 quasars selected from SDSS Data Release 5 \rev{\citep{2007AJ....134..102S}} with $g < 18.5$ mag, $2.3 < z_{\rm em} < 2.6$, observed with the WFC3/UVIS-G280 grism in Cycle 17.
These data were taken specifically for the scientific goal of surveying the abundance of strong \ion{H}{i} Lyman limit absorption features at $1.2 < z < 2.5$.
Flux and wavelength calibrated 1D spectra for each quasar in this sample are extracted using customized software (see Section 3.1 in O11).
WFC3/UVIS-G280 spectra span roughly $\lambda=2000-6000$ \AA~and they have
relatively high signal-to-noise ratio (S/N$\sim$20) per pixel down to $\lambda \sim 2000$\AA\footnote{The data generally have S/N exceeding 10 pixel$^{-1}$ at all wavelengths $\lambda >$ 2000\AA.}~($FWHM\sim60$\AA~at $\lambda=2500$\AA).
Uncertainties in the wavelength calibration are of the order of 2 pixels, in the form of a rigid shift in the pixel space (see Table 1 in O11). 

As already pointed out by \citet{2013ApJ...765..137O} (O13 hereafter), since this sample is based on SDSS optical colour selection \citep{2002AJ....123.2945R}, it might be biased towards bluer colours than a complete sample. However, \citet{2011ApJ...728...23W} have demonstrated that at the redshifts of this sample this bias is relatively small (see their Figure 16).

Note however that quasar samples constructed from UV spectroscopy
archives are based both on optical colour-selection and additional UV
brightness criteria. Typically the brightest objects observed in previous \hst cycles were 
re-observed when a new UV instrument came online. Objects observed by the \hst Cosmic Origins
Spectrograph were observed in the post GALEX era, and are thus explicitly biased to have bright
near-UV and far-UV magnitudes. The selection function of quasars in the UV archives is thus unknown and extremely difficult to quantify, but the expectation is that such samples are biased very  blue. In contrast, the selection criteria for our quasar sample is much cleaner, as it is simply an optical apparent magnitude limited sample $g < 18.5$ of quasars at  $z_{\rm em}\simeq 2.4$. 

\subsection{X-rays}
\label{X-rays}

The X--ray data have been gathered from the ROSAT, XMM-{\it Newton}, and {\it Chandra} archives. 
We found 2 objects (J1253+0516 and J1335+4542) detected in the ROSAT All-Sky Survey Faint Source Catalog (RASS-FSC), 6 quasars have at least one ROSAT/PSPC \citep{1994AAS...185.4111W,1994IAUC.6100....1W} observation, while 2 sources have ROSAT/HRI images (J1454+0325 and J1119+1302 with also a PSPC observations).

The X--ray fluxes at 0.5--2 keV were calculated from the count rates in the ROSAT band (0.1--2.4 keV) and in the PSPC band  (0.24--2 keV) by utilising a power law spectrum with a photon index $\Gamma=2.0$ modified by Galactic absorption \citep{2005A&A...440..775K} only. The X-ray properties of the objects in our sample detected by ROSAT are listed in Table~\ref{tbl-x}.

Sources without a ROSAT detection might be either intrinsically faint, X--ray obscured, and/or highly variable. We expect the undetected AGN to have X--ray luminosities of the order of $7\times10^{45}$ erg s$^{-1}$ or lower in the rest-frame 0.5--2 keV band\footnote{The ROSAT RASS flux limit is $5\times10^{-13}$ erg s$^{-1}$ cm$^{-2}$ for a mean effective exposure time of 400 sec and $\Gamma=2.0$, but the range of ROSAT exposure times is large, and the sensitivity limit is different from field to field (\citealt{1999A&A...349..389V}).}.

Among these 10 quasars, two have additional spectral information from XMM-{\it Newton} (J0755+2204 and J1119+1302) and one from {\it Chandra} (J1220+4608). 
The soft ($S=0.5-2$ keV) and hard ($H=2-10$ keV) X-ray luminosities for J0755+2204 and J1119+1302 are tabulated in the SDSS (DR5)/XMM--Newton quasar survey catalog, and the values are $L_{\rm S}=4.60\times10^{44}$ and $L_{\rm H}=1.14\times10^{45}$ erg s$^{-1}$ for J0755$+$2204, while J1119$+$1302 has $L_{\rm S}=1.96\times10^{45}$ and $L_{\rm H}=5.00\times10^{45}$ erg s$^{-1}$.

J1220$+$4608 was targeted by {\it Chandra} with ACIS-S in March 2008 with a 
nominal exposure of 50 ks. The spectrum has very low number counts, with a $\leq3\sigma$ detection in the soft band. 
In this case, assuming a photon index of 1.8 and Galactic $N_{H}$, we found that the extrapolated flux in the 2--10 keV band is $1.36\times10^{-14}$ erg s$^{-1}$ cm$^{-2}$. This flux corresponds to a luminosity of $6.4\times10^{44}$ erg s$^{-1}$.

Summarising, the fraction of detected sources is 21\% (i.e. 11/53, 10 objects detected by ROSAT/XMM--Newton and one additional source with a {\it Chandra} observation) with a mean soft X--ray luminosity of about $5.6\times10^{45}$ erg s$^{-1}$.

\subsection{Radio}
\label{Radio}

To estimate the fraction of radio emitters in our sample, we matched it with the 
DR7 quasar property catalog by \citet{2011ApJS..194...45S}. 
Radio properties are collected from FIRST \rev{\citep{1995ApJ...450..559B}} and from NRAO/VLA Sky Survey \citep{1998AJ....115.1693C} with a matching radius of 30" (our matches are all within 1.3''). We have found 11 quasars in the FIRST survey, while one object (J2338+1504) has a detection from NRAO, for a total of 12 detections.
The radio loudness is estimated following \citet{2007ApJ...656..680J}, $R = f_{6{\rm cm}} /f_{2500}$ where $f_{6{\rm cm}}$ and $f_{2500}$ are the flux density at rest-frame 6 cm and 2500 \AA, respectively. 
According to Jiang et al., 11 quasars are core-dominant and one is lobe-dominant (J1119+1302, radio morphology classification following \citealt{2007ApJ...656..680J}). 
Five objects are not in the FIRST footprint (flag = -1 in the DR7 catalog, J1259+6720, J1325+6634, J1400+6430, J2111+0024, J2136+1029) and they do not have any detection from NRAO.
A summary of the radio properties is given in Table~\ref{tbl-r}.
Following \citet{kellermann89}, quasars with $R > 10$ are defined as ``radio-loud".
In the WFC3 sample the radio-loud fraction (RLF) is of the order of 19\%. This is consistent with the RLF of quasar at similar optical magnitudes \footnote{\citet{2007ApJ...656..680J} have found that the RLF is about 15\% for an absolute magnitude at the rest-frame 2500\AA~($M_{2500}$) of $\sim-$28, which is consistent with the average $M_{2500}$ of the WFC3 sample ($M_{2500}\simeq-$29).}.

In the following, we will present the WFC3 average spectrum with and without the 10 objects with $R>10$ (a stand alone radio-loud quasar stack is not constructed given the poor statistics).

\subsection{GALEX}
\label{GALEX}

To extend our wavelength coverage to shorter wavelengths, we first considered the GALEX photometry in 
the DR7 quasar property catalog. We found that the detection rate for the near-UV (NUV at
$\lambda_{\rm eff}=2316$\AA) and far-UV (FUV at $\lambda_{\rm eff}=1539$\AA) bands was 64\% (34/53) and 14\% (14/53), respectively. Given the low detection rate, we have cross matched our sample to the GALEX forced photometry catalog (David
Schiminovich, private communication), which allows us to obtain a detection for almost all sources in the WFC3 sample. 
The GALEX forced photometry was not available for only 4 objects (J0751+4245, J1354+5421, J1540+4138,
J2111+0024), for the rest of the sample we have both NUV and FUV bands. 
The observed GALEX forced photometry for the WFC3 sample is listed in Table~\ref{tbl-g}.
The average GALEX bands are then corrected for IGM absorption and we will outline how these data have been corrected for IGM absorption in Section~\ref{Comparison with GALEX}.

\begin{figure}
 \includegraphics[width=\linewidth,clip]{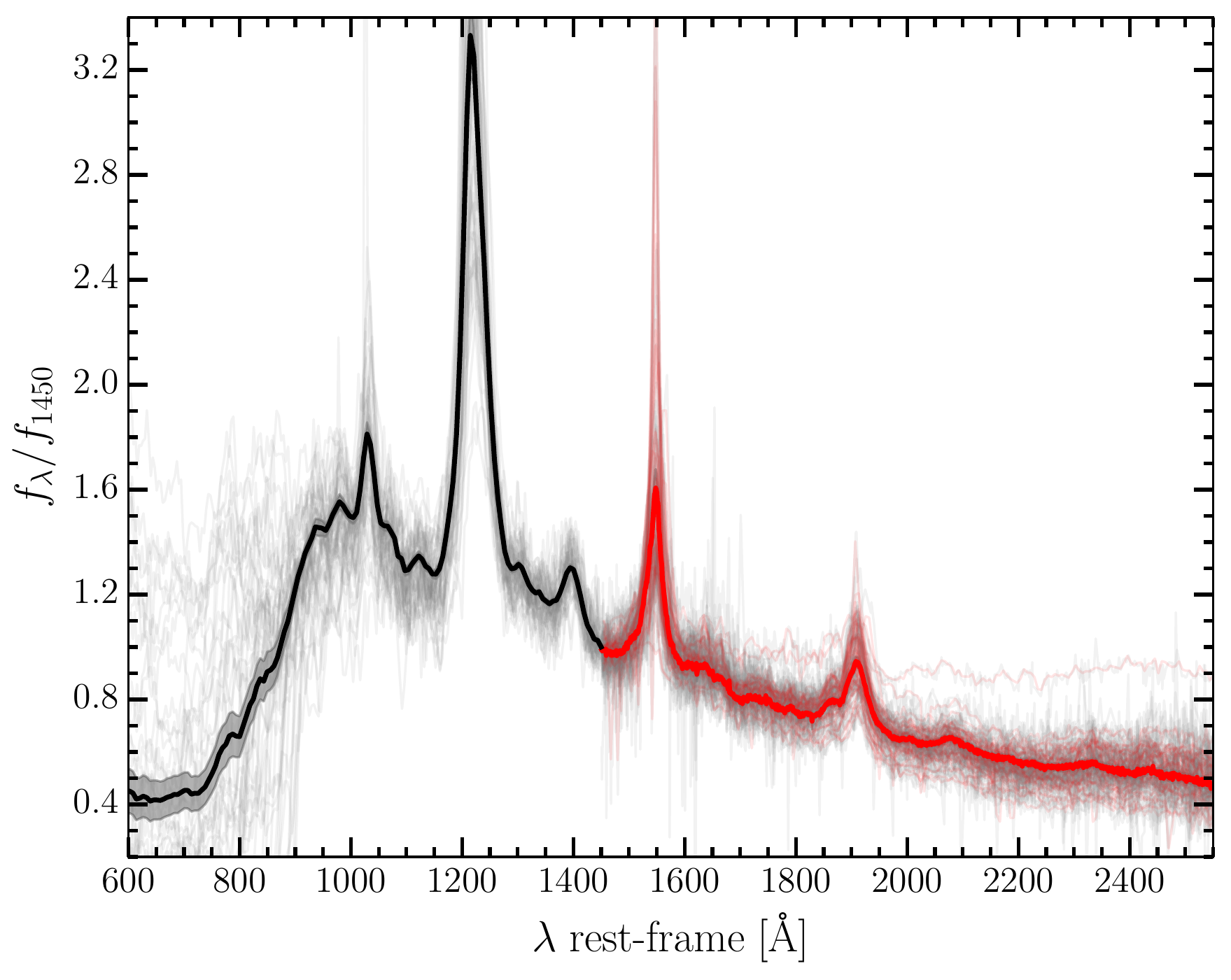}
 \caption{ \rev{Mean observed QSO} spectrum as estimated from a stack of the WFC3 spectra, each normalized to unit flux at 1450\AA  (see \S\ref{Composite construction} for details). The black line represents the WFC3 QSO average spectrum with uncertainties from bootstrap (black shaded area). The red line represents the SDSS QSO average spectrum (not smoothed to the WFC3 resolution) of our WFC3 sample with uncertainties. Grey lines represent the single QSO spectra, while thin red lines represent SDSS spectra smoothed to WFC3 resolution.}
 \label{wfc3sdss_stack}
\end{figure}

\section{Average spectrum construction}
\label{Composite construction}

We follow a similar procedure as O13 for the construction of the WFC3 average spectrum. Specifically:
\begin{enumerate}
 \item We correct the quasar flux \rev{density\footnote{\rev{In the following we will use the word ``flux" to mean the flux density (i.e. flux per unit wavelength).}} ($f_\lambda$)} for Galactic reddening by adopting the \ebv estimates from \citet[SFD, median reddening value is \ebv$=0.02$ mag]{schlegel98} and the Galactic extinction curve from \citet{1999PASP..111...63F} with $R_V=3.0$. The same reddening law has been considered to correct the GALEX fluxes.

 \item We generate a rest-frame wavelength array with fixed dispersion $\Delta\lambda$. The dispersion value was set to be large enough to include at least one entire pixel from the WFC3/UVIS-G280 spectra at rest wavelengths $\lambda < 1215$\AA~(i.e. $\Delta\lambda\simeq6.2$\AA).

 \item Each quasar spectrum was shifted to the rest-frame and linearly interpolated over the rest-frame wavelength array with fixed dispersion $\Delta\lambda$\footnote{Wavelengths are divided by $(1+z)$ to shift the spectra into the source rest frame, while fluxes in $f_\lambda$ are multiplied by $(1+z)$.}.

 \item We normalized single spectra by their flux at rest $\lambda =1450$\AA.

 \item All the flux values were then averaged to produce the stacked spectrum normalized to unity at $\lambda=1450$\AA.  
\end{enumerate}
\rev{Recently, \citet{2013ApJ...771...68P} have found that UV colours at high latitudes are best fitted with a \citet{1999PASP..111...63F} extinction curve with $R_V\simeq2.2$, while \citet{2011ApJ...737..103S} found that SFD overestimates reddening by 14\% (reddening values should be recalibrated as $E(B-V) = 0.86 \times E(B-V)_{SFD}$). 
The average WFC3 spectrum re-estimated following the results outlined above is fully consistent with the one constructed with $R_V=3.0$ and $E(B-V)$ from SFD.}

Uncertainties on the observed stack are estimated through the bootstrap resampling 
technique.  We created 5000 random samplings of the 53 spectra with replacement, and we applied the same procedure as described above. The resulting stack is shown as the solid black line in Figure~\ref{wfc3sdss_stack} for the full sample, while the resulting uncertainties on the stacked spectrum are plotted with a shaded area.
The same technique is applied to the SDSS quasar spectra and the resulting stacked spectrum is presented in Figure~\ref{wfc3sdss_stack} with the red solid line. For plotting purpose we show the SDSS stacked spectrum down to $\lambda=1450$\AA~given that the WFC3 stack already \rev{covers wavelengths below this region}.

To bring the WFC3 and the SDSS stacked spectra over a common luminosity scale in the $\nu L_\nu$ plane, we have multiplied the final stacks for the mean flux (estimated from the WFC3 and SDSS) of the sample at 1450\AA. For the sake of matching the flux scale we have convolved the SDSS spectra to WFC3 resolution. 
The mismatch between the WFC3 and SDSS spectra at 1450\AA~is 11\% and it is due to variability and flux calibration errors \rev{in the WFC3 data (see Section 3.2 in O11 for details)}. We have thus re-scaled the WFC3 stacked spectrum to match the (convolved) SDSS at 1450\AA.

\begin{figure}
 \includegraphics[width=\linewidth,clip]{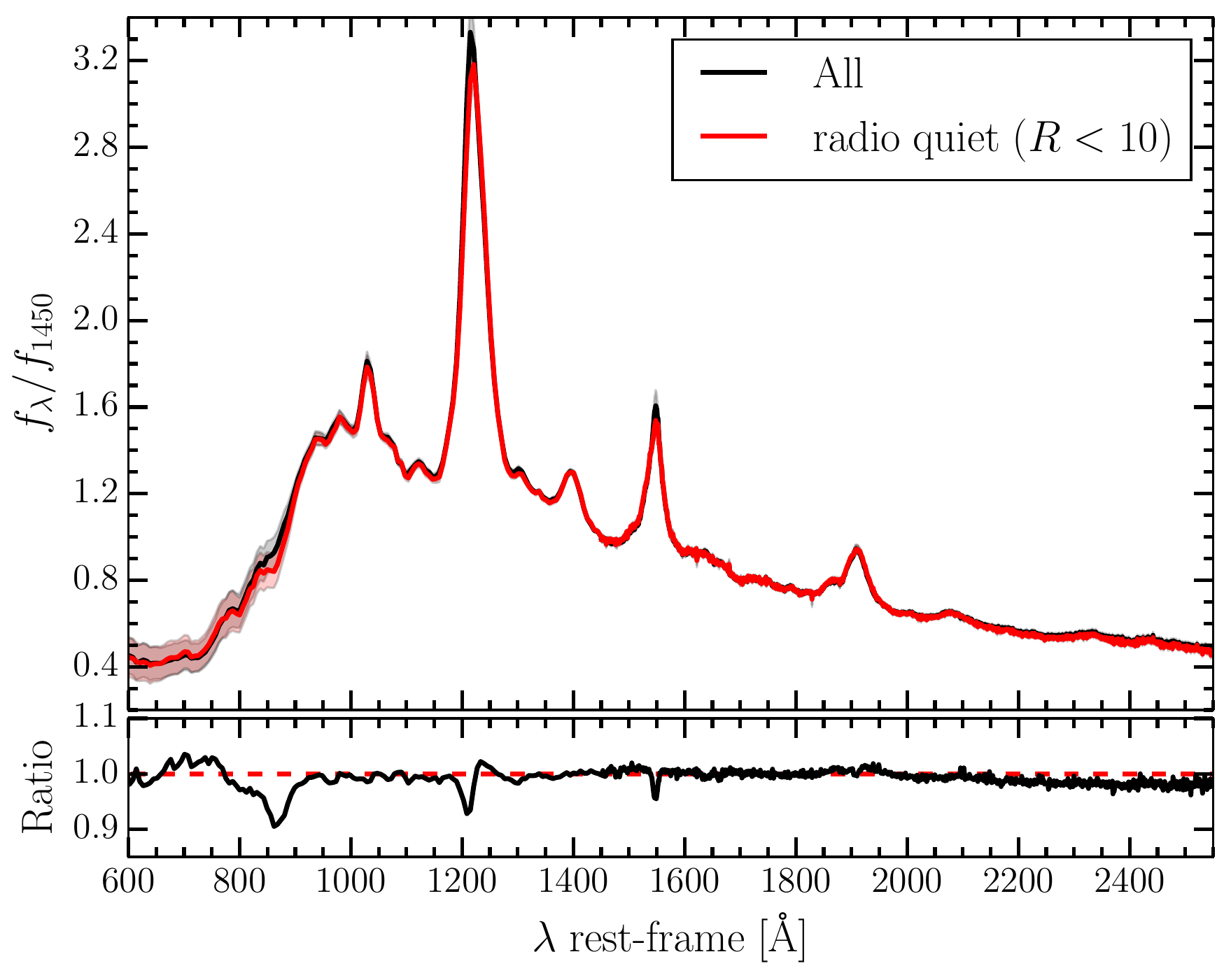}
 \caption{Upper panel: \rev{Mean observed QSO} spectrum for the WFC3 sample (black line) and for the subsample with $R<10$ (red line). each normalized to unit flux at 1450\AA. \rev{Lower Panel: Ratio of the $R<10$ to the full mean observed QSO spectrum.}}
 \label{all_rq}
\end{figure}
We have also constructed the quasar average spectrum by excluding the 11 objects with $R>10$ for completeness. The resulting spectrum is shown in Figure~\ref{all_rq} where we have over plotted the WFC3 average spectrum as a comparison. The radio quiet quasar stack is fully consistent with the WFC3 stack within the uncertainties.

\section{IGM transmission correction}
\label{IGM transmission correction}
\begin{figure}
 \includegraphics[width=\linewidth,clip]{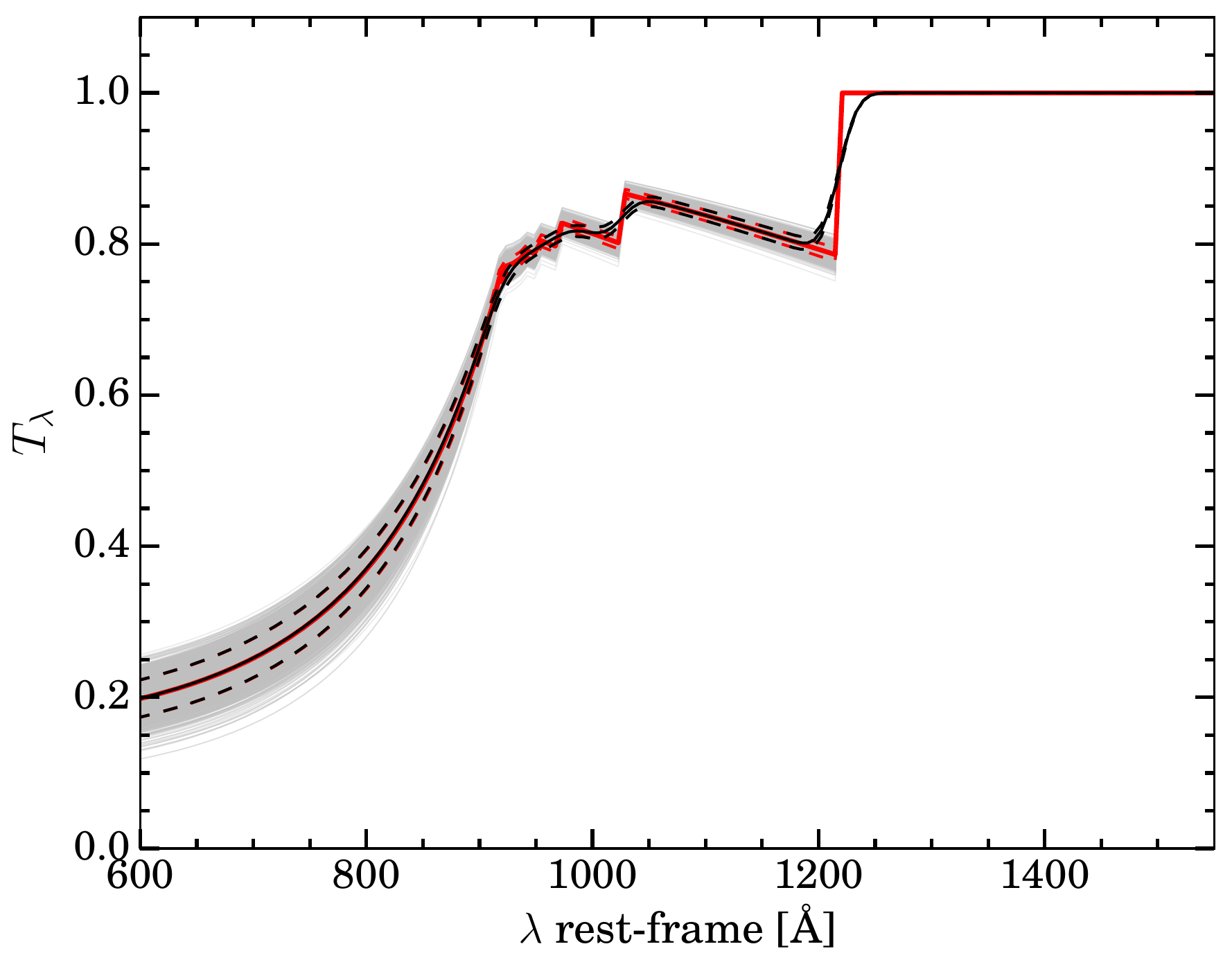}
 \caption{IGM transmission curves as a function of rest-frame wavelength. Grey curves represent 10,000 different realisations of $\Tl$, while the solid and dashed red curves are the stack and $1\sigma$ dispersion, respectively. The black solid curve is the smoothed stack to the WFC3 resolution with the corresponding $1\sigma$ dispersion (black dashed lines).}
 \label{igmfct}
\end{figure}

Blueward of Lyman alpha emission in the quasar rest frame, absorption from
intergalactic \ion{H}{i} attenuates the quasar flux, both in the Lyman series
(creating the so-called \ion{H}{i} forest), and in the Lyman continuum at rest
$\lambda<912$\,\AA\ \citep[e.g.][]{1990A&A...228..299M}.
The large abundance of neutral gas at $z_{\rm em}\la 2.4$ is very apparent in our average quasar spectrum shown in Fig.~\ref{wfc3sdss_stack}. 
With only mild assumptions
on the average quasar SED, the exponential flux decline at
$\lambda<912$\,\AA\ yields the mean free path of Lyman limit
photons in the IGM (O13). Here we reverse the question and constrain the
quasar SED for a range of IGM transmission curves $\Tl$ for
$\lambda<1215.67$\,\AA.

For a source at emission redshift $z_\mathrm{em}$ the effective optical depth
to \ion{H}{i} Lyman series and Lyman continuum photons at redshift $z<z_\mathrm{em}$
is determined from the \ion{H}{i} absorber distribution function in redshift and
column density $\fnz=\partial^2n/\left(\partial N_\mathrm{HI}\partial z\right)$. 
The resulting average IGM transmission $\Tl$ critically depends on the parametrization
of $\fnz$ \citep{1995ApJ...441...18M,2006MNRAS.365..807M,2014MNRAS.442.1805I} 
and is statistical in nature due to
the stochasticity of Lyman limit systems \citep{1999ApJ...518..103B,2008MNRAS.387.1681I,2011ApJ...728...23W}.
Quasar composites based on low-resolution spectra need to be corrected
for Lyman series and Lyman continuum absorption of low-column density absorbers that
cannot be identified and corrected by eye. 
The statistical IGM correction strongly depends on the assumed absorber distribution
parameters and their dependence on redshift, which may have resulted in large
systematic errors in existing quasar composite spectra if the incorrect parameters were
used or if redshift evolution was not properly taken into account. 
Moreover, the ability to identify weak partial Lyman limit systems
($\tau_{912}\la 0.3$) depends on redshift, the employed spectra (S/N,
spectral resolution, wavelength coverage, flux calibration), and the
intrinsic quasar continuum, such that heterogeneous samples are prone
to ambiguities in the continuum definition and the treatment of Lyman
limit systems.  In particular, we stress that applying an average
correction for partial Lyman limit systems to de-attenuate individual
$z_\mathrm{em}\la 3$ sightlines is incorrect because of the
stochasticity of Lyman limit absorption\citep{2011ApJ...728...23W}.

Our large sample at similar $z_\mathrm{em}\sim 2.4$ ensures a
sufficient sampling of partial Lyman limit systems, justifying a
redshift-specific IGM correction function $\Tl$ to be applied to our
stacked spectrum (Fig.~\ref{wfc3sdss_stack}).  However, the low
spectral resolution prevents an unambiguous identification of weak
partial Lyman limit systems in individual spectra without knowledge of
the underlying quasar continuum.  We therefore have to correct for the
average Lyman series and continuum absorption of the whole \ion{H}{i}
absorber population statistically. While this is the simplest and most
liberal approach, the significant flux drop in the Lyman continuum
implies a large correction to our stacked spectrum, with additional
uncertainties related to the parametrization of the IGM.  

In this analysis we consider the recent constraints on the $z\sim 2.4$ IGM
presented by \citet[and references therein]{2014MNRAS.438..476P} within their range of uncertainty.
\citet{2014MNRAS.438..476P} introduced a cubic Hermite spline model to describe $\fnz$. They performed a Monte Carlo Markov Chain (MCMC) analysis of existing
constraints on $\fnz$ to derive the posterior probability distribution
functions of seven spline points spaced at irregular logarithmic
intervals in the range $\mnhi = 10^{12}$--$10^{22}$\,cm$^{-2}$.  Using
the output from their MCMC chains\footnote{http://www.arcetri.astro.it/$\sim$lusso/Site/Research.html}, we generated 10,000 realizations of $\fnz$ at $z=2.4$ and calculated
$\Tl$ in the observed wavelength range with a semi-analytic technique.
This modeling assumes that the \ion{H}{i} forest is composed of
discrete ``lines'' with Doppler parameter $b = 24\,\mkms$ and that the
normalization of $\fnz$ evolves as $\left(1+z\right)^{0.5}$ which
implies increasing transmission below
$\lambda=912$\,\AA\ for the lower-redshift Lyman series.
Opacity due to metal line transitions was ignored since they
contribute negligibly to the total absorption in the Lyman continuum.
In Fig.~\ref{igmfct} we plot our suite of 10,000 IGM transmission
functions with grey lines.  We then constructed the mean of all of
these different realizations and smoothed this to the WFC3 grism
resolution (5 pixels).  Finally, we shifted the observed wavelengths
to rest frame at $z=2.4$ and resampled the transmission functions onto
the rest-frame wavelength grid of our stacked quasar spectrum. 
This defines our mean IGM transmission function which is shown as 
the black curve in Figure~\ref{igmfct}. 
Given the narrow redshift range of our quasar sample, we did not
account for redshift evolution in $\Tl$. To summarize, by treating
$\fnz$ within reasonable uncertainties our approach yields a range of
plausible IGM correction functions, a clear advance over previous
analyses that assumed a fixed correction.  However, since we account
for the total \ion{H}{i} absorber population, our IGM correction
functions in Fig.~\ref{igmfct} assume that the column density
distribution is well sampled at high column densities.

\begin{figure}
 \includegraphics[width=\linewidth,clip]{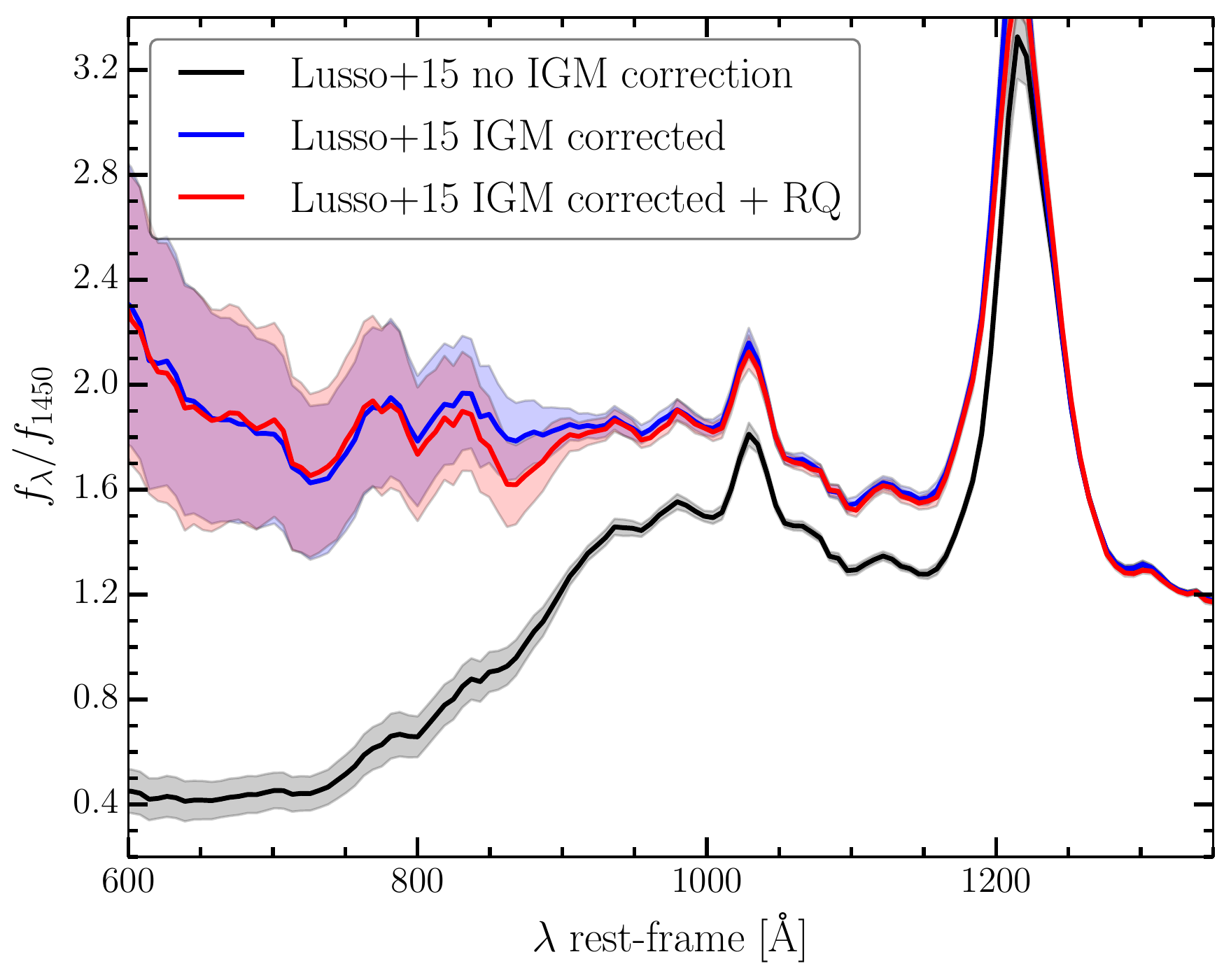}
 \caption{Mean IGM corrected QSO spectrum with uncertainties from bootstrap (shaded area) for the full WFC3 (blue) and for the radio quiet (red) sample. The black line represents the observed WFC3 QSO stack with uncertainties from bootstrap (grey shaded area).}
 \label{igmfctcorr}
\end{figure}

The procedure to correct the observed WFC3 spectra for IGM absorption is outlined as follows:
\begin{enumerate}
 \item We generate a set of 10,000 mock quasar stacks, following the same 
 procedure as in Section~\ref{Composite construction}, by drawing randomly from the 53 quasar
spectra to assess sample variance allowing for duplications. 
 \item We then randomly draw one IGM transmission
function from our suite of 10,000. We smoothed this to the WFC3 grism
resolution (5 pixels), and we resampled the transmission function onto
the rest-frame wavelength grid of our stacked quasar spectrum. This  
is repeated for each mock quasar stack.
 \item We divide the observed spectral flux ($f_{\lambda,\rm obs}$) by the IGM
transmission curve 
\begin{equation}
	f_{\lambda,{\rm corr}}=f_{\lambda,{\rm obs}}/T_{\lambda}.
\end{equation}
 \item The 10,000 mock stacks corrected from IGM absorption are then averaged to produce the stacked spectrum (normalized to unity at $\lambda=1450$\AA).
 \item The uncertainties on the corrected WFC3 stacked spectrum are estimated from the dispersion of these 10,000 mock stacks.
\end{enumerate}
The resulting stacks for the full WFC3 and the radio quiet samples
are shown in Figure~\ref{igmfctcorr} with the blue and red lines,
respectively, and are tabulated in Table~\ref{tab:spectra}.  The
stacked spectra show a softening at wavelengths $\lambda<912$\AA\ and
several emission lines. We will discuss these results further in
Section~\ref{Results}.  The 1$-\sigma$ uncertainties on the constructed
stack are displayed as a light blue shaded area. 

One possible concern about the measure of the uncertainties on the stack is that 
the bootstrap technique outlined above may not have converged far in the blue.
In the limit where only a handful of the same transmitting spectra actually contribute to the stack at $\lambda<912$\AA\ the bootstrap may underestimate the true noise in our measurement.
In general, the overall uncertainty on the stack depends on a
variety of effects such as the intrinsic fluctuations of the
underlying quasar continuum (sample variance), our imperfect knowledge
of the IGM transmission function (due to uncertainties in the $\fnz$;
see Figure~\ref{igmfctcorr}), spectral noise (i.e. read noise and photon counting
noise), and most importantly, Poisson shot-noise (sample
variance) due to the number of LLS and partial LLS intercepted by the
quasar sightline. 

As we have discussed above, our suite of 10,000 $\Tl$ are created from an integral over the column
density distribution, which includes uncertainties in the parameters governing the $\fnz$, but it 
does not include the variance due to the stochasticity of LLS absorption.

To this end we conducted a Monte-Carlo simulation with \rev{a different set of empirical IGM transmission curves following the IGM parametrization described in O13}. 
We assumed the $\fnz$ defined in Table~10 in O13 with a $(1+z)^{1.5}$ redshift evolution in the normalization
(see their Equation 6). For each quasar in our sample we drew 100 sightlines from a parent
sample of 5000 sightlines populated with $0<z<3$ absorbers and computed mock WFC3 IGM transmission spectra in the covered wavelength range, i.e.\ at
2000\,\AA$<\lambda<1215.67\left(1+z_\mathrm{em}\right)$\,\AA.
This allows for the small redshift evolution in the IGM transmission in the narrow redshift
range covered by our sample. The result is a set of 100 mock spectra per quasar that accounts
for the Poisson statistics of Lyman limit systems. This method does not consider uncertainties
in the shape of $\fnz$, but samples $\fnz$ properly at high column densities
(albeit without clustering). 
The Monte-Carlo simulation is then carried out as below:
\begin{enumerate}
  \item We start by assuming the true underlying mean QSO spectrum to be the one corrected by the $\Tl$ functions.
  \item We then create a large number (i.e. 8000) of mock samples of 53 quasars by drawing randomly from the 100 IGM transmission realisations for each quasar in the mock sample.  
  \item We finally calculate the variance of this stack from these many ensembles of 53.
\end{enumerate}
This procedure fully encapsulates the stochastic nature of IGM absorption, and in particular shot noise
due to the presence or absence of LLSs.
\begin{figure}
 \centering\includegraphics[width=\linewidth,clip]{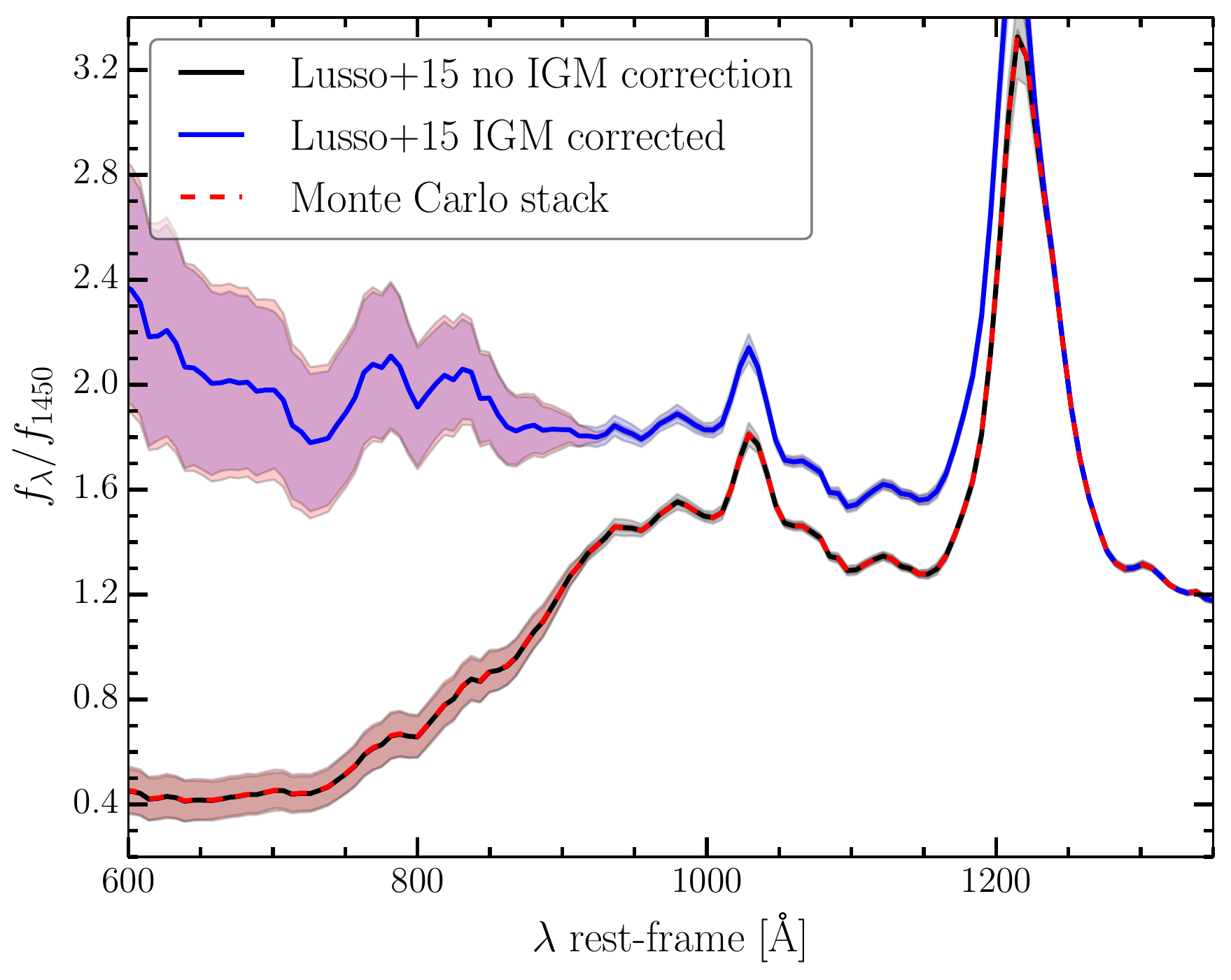}
 \caption{Comparison between the Monte Carlo stack and $1-\sigma$ uncertainties described in \S~\ref{IGM transmission correction} ({\it red dashed line}) with the one estimated by applying the bootstrap technique on the data. The red shaded area matches the uncertainties due data bootstrap, highlighting the fact that the bootstrapping is converged far in the blue. }
 \label{shotnoise}
\end{figure}
Figure~\ref{shotnoise} presents the comparison between the uncertainties from bootstrap and those estimated following the above approach. The amplitude of the uncertainty on the Monte Carlo stack matches the one estimated by applying the bootstrap technique on the data. Our Monte Carlo simulations suggest a $\sim$16\% fluctuation in samples of 53 due solely to shot-noise in the IGM absorption. Since this fluctuation is comparable to the bootstrap error in our observations, this strongly suggests that the IGM fluctuations dominate the error budget in the stack, and that the other sources of variance (i.e. intrinsic variations in the quasar continuum and spectra noise) are sub-dominant.
Most importantly, this comparison also shows that the bootstrapping is converged far in the blue. In fact, given the narrow AGN redshift range, which goes from $z_{\rm min}=2.282$ to $z_{\rm min}=2.599$ (mean redshift $\langle z_{\rm em} \rangle\sim2.44$), almost all spectra contribute appreciably to the total flux at 600 \AA. 
\rev{We note that the stack estimated with this second set of IGM transmission curves is in agreement, within the uncertainties, with the one constructed from the 10,000 $\Tl$ created from an integral over the column density distribution. We have thus decided to consider the stack constructed from the 10,000 IGM transmission curves for the rest of our analysis.}
 
%
\begin{table}
  \caption{WFC3/UVIS Stacked Spectrum corrected for IGM absorption}
  \label{tab:spectra}
  \begin{center}
    \begin{tabular}{ccccccc} \hline \hline  
                         &  \qquad\qquad \qquad All  &    & \qquad\qquad RQ &   &  \\ 
    $\lambda^{\mathrm{a}}$   &   $\flf^{\mathrm{b}}$  & $\sigma(\flf)^{\mathrm{c}}$ & $\flf$  & $\sigma(\flf)$ \\ \hline 
   583.351   &   2.382   &   0.577   &   2.405     &   0.602  \\
   589.540   &   2.247   &   0.548   &   2.302     &   0.580  \\
   595.729   &   2.326   &   0.542   &   2.345     &   0.576  \\
   601.919   &   2.297   &   0.533   &   2.248     &   0.551  \\
   608.108   &   2.234   &   0.518   &   2.205     &   0.548  \\
   614.297   &   2.092   &   0.492   &   2.107     &   0.522  \\
   620.486   &   2.081   &   0.472   &   2.049     &   0.492  \\
   626.675   &   2.090   &   0.474   &   2.044     &   0.495  \\
   632.865   &   2.037   &   0.462   &   1.996     &   0.477  \\
   639.054   &   1.945   &   0.444   &   1.911     &   0.467  \\
   645.243   &   1.937   &   0.428   &   1.916     &   0.448  \\
   651.432   &   1.908   &   0.418   &   1.888     &   0.442  \\
   657.621   &   1.872   &   0.402   &   1.864     &   0.423  \\
   663.811   &   1.866   &   0.388   &   1.871     &   0.413  \\
   670.000   &   1.867   &   0.388   &   1.893     &   0.413  \\
   676.189   &   1.850   &   0.379   &   1.890     &   0.405  \\
   682.378   &   1.848   &   0.373   &   1.857     &   0.394  \\
   688.567   &   1.814   &   0.363   &   1.831     &   0.384  \\
   694.757   &   1.814   &   0.354   &   1.847     &   0.375  \\
   700.946   &   1.811   &   0.340   &   1.866     &   0.370  \\
   \hline
    \end{tabular}
    
 \flushleft\begin{list}{}
 \item {\bf Notes.}
 \item${}^{\mathrm{a}}${ Rest-frame wavelength in Angstrom.}
 \item[]${}^{\mathrm{b}}${ Mean IGM corrected flux per \AA~normalized to the flux at 1450\AA.}
 \item[]${}^{\mathrm{c}}${ Flux uncertainties from our bootstrap analysis (see text).}
 \end{list}
 \vspace{0.2cm}
 (This table is available in its entirety in a machine-readable form in the online journal. A portion is shown here for guidance regarding its form and content.)
  \end{center}
\end{table}

\section{Results}
\label{Results}

%
\begin{table}
  \caption{Emission lines properties}
  \label{tab:emlines}
  \begin{center}
    \leavevmode
    \begin{tabular}{lccc} \hline \hline              
    Line  & $\lambda_{\rm obs}$   & Flux$^{\mathrm{a}}$  &   EW \\  
      &  (\AA)   & (erg s$^{-1}$ cm$^{-2}$ \AA$^{-1}$)  &   (\AA) \\ \hline 
  \ion{H}{i}-\ion{Ly}{$\alpha$}+\ion{S}{iv} & 938.0 & $8.2\pm2.7\times10^{-16}$ & 0.5  \\
  \ion{Ly}{$\gamma$}+\ion{C}{iii}]            & 987.6  & $4.4\pm1.1\times10^{-15}$  & 3.0  \\
  \ion{Ly}{$\beta$}+\ion{O}{iv}                  & 1029.7  & $9.8\pm0.4\times10^{-15}$  &  8.5  \\
  \ion{Fe}{ii}+\ion{Fe}{iii}                          & 1122.0  & $1.8\pm1.0\times10^{-15}$  &  1.3  \\
  \ion{Ly}{$\alpha$}                                  & 1216.1  &  $9.0\pm0.9\times10^{-14}$ &  74.0  \\
  \ion{Si}{iv}+\ion{O}{iv}]                          & 1397.5  & $6.7\pm1.4\times10^{-15}$  & 7.2   \\
  \ion{C}{iv}                                             & 1544.6   &  $1.7\pm0.2\times10^{-14}$ & 17.8   \\
  \ion{He}{ii}                                            & 1635.7  & $1.2\pm0.3\times10^{-15}$  & 2.0   \\
  \ion{O}{iii}]                                            & 1663.7  &  $7.3\pm1.0\times10^{-16}$ &  1.2  \\
  \ion{Al}{iii}                                             & 1861.0 &  $1.8\pm0.5\times10^{-15}$ &  2.6  \\
  \ion{C}{iii}]                                            & 1907.3 & $9.7\pm0.9\times10^{-15}$  &  15.5  \\
   \hline
    \end{tabular}
 \flushleft\begin{list}{}
 \item {\bf Notes.}
 \item${}^{\mathrm{a}}${ Normalized fluxes are multiply by the average flux of the WFC3 sample at 1450\AA.}
 \end{list}  \end{center}
\end{table}
\begin{figure*}
 \includegraphics[width=0.52\textwidth]{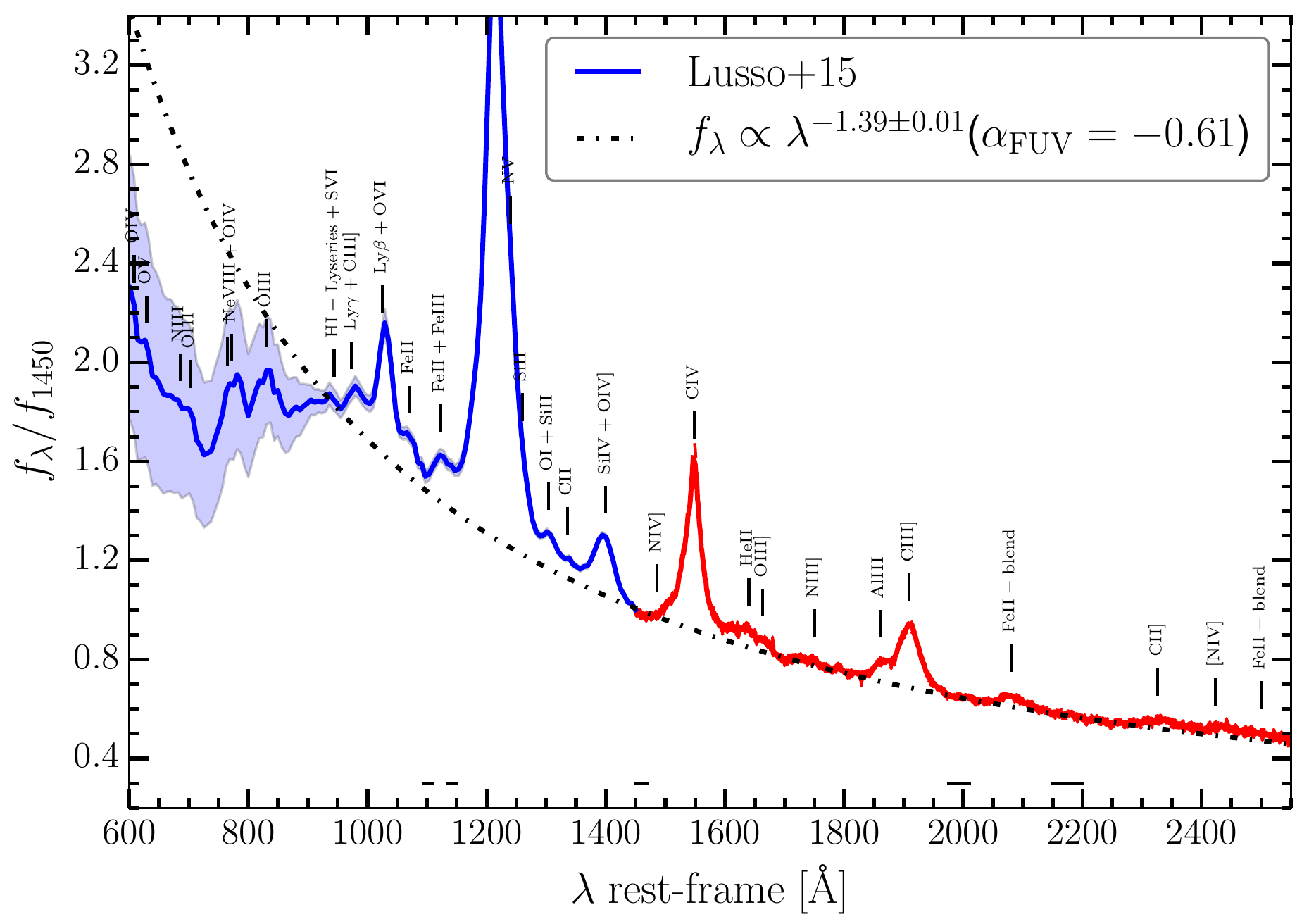}
 \includegraphics[width=0.47\textwidth]{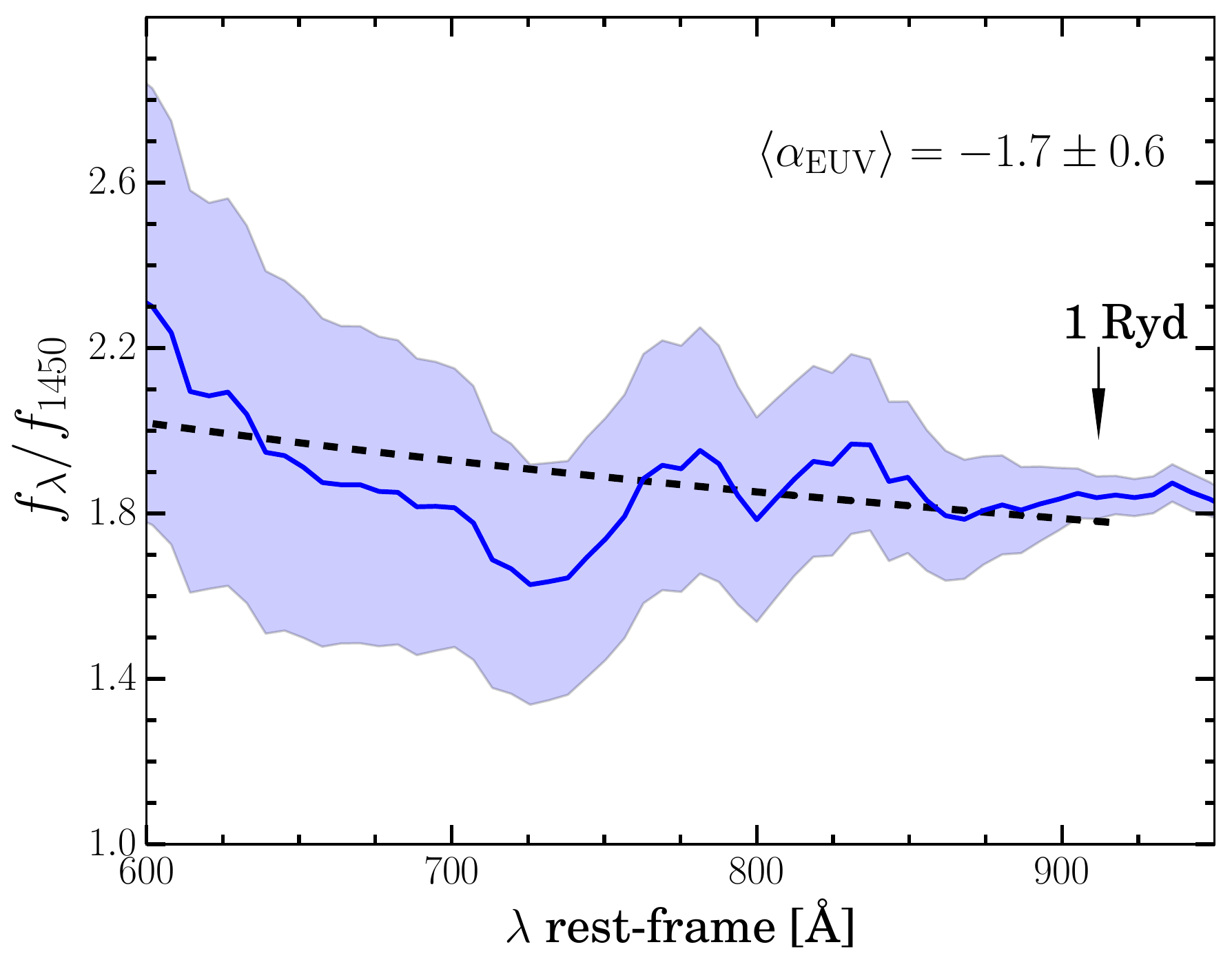}
 \caption{Left panel: Average WFC3 and SDSS spectra normalized at 1450\AA\ with fits to power-law continuum (dot-dashed line). Five continuum windows (from T02) are shown as black horizontal lines below the spectrum. The power-law continuum fit exhibits a break at $\lambda\simeq920$\AA, with a flatter (softer) spectrum at shorter wavelength. Right panel: Zoom in of the EUV region (600$-$912\AA). The dashed line is the power-law continuum by employing the average EUV spectral slope.}
 \label{fit}
\end{figure*}
\subsection{Spectral fit and emission lines}
\label{Spectral fit and emission lines}

The AGN ionising continuum and its shape in the optical-UV is crucial for several reasons, such as interpreting broad and narrow emission-lines observed in AGN spectra, and their relative ratios. It also defines the BBB (i.e. the bulk of the QSO emission), which can be important in interpreting the observed relation between optical and soft X--ray fluxes.

In order to infer the continuum slope of the WFC3 stack, we have to avoid contamination of broad emission lines, broad absorption features and extended wings of emission lines\footnote{Emission from host-galaxy star light is not considered in the fit since our stack covers the rest-frame UV range of very bright AGN, where the expected contribution of the galaxy should be negligible.} especially at $\lambda>1216$\AA\ where the emission lines are more prominent. 
 
We have thus fitted the UV stack in each wavelength window free of strong features by means of a single power law. We adopted the same intervals as T02, but given that the uncertainties dramatically increase at $\lambda<912$\AA\, we decided to restrict the fit of the continuum redwards of 912\AA\ (i.e. 1095$-$1110, 1135$-$1150, 1450$-$1470, 1975$-$2010, and 2150$-$2200 \AA). 
We note that the continuum window blueward of \ion{Ly}{$\alpha$} (i.e. 1095$-$1110\AA) may be contaminated by the \ion{Fe}{ii}+\ion{Fe}{iii} emission line. Given our limited resolution we decided to keep it for consistency with T02. Results are unaffected if we neglect this window.
We have considered as the error for the $\chi^2$ minimization procedure the uncertainties in the stack, which are rather uniform at \rev{$\lambda>912$ \AA}.
Figure~\ref{fit} shows the rest-frame stacked spectrum extending from 600 \AA~to 2500 \AA~and the power-law fit to the continuum of the form $f_\lambda\propto\lambda^{\alpha_\lambda}$, where the best-fit power law index is $\alpha_\nu=-0.61\pm0.01$\footnote{In the following we will refer to $\alpha_\nu$ only. The relation between the fluxes in wavelength, $f_\lambda\propto\lambda^{\alpha_\lambda}$, and frequencies, $f_\nu\propto\nu^{\alpha_\nu}$, is $\alpha_\nu = - (2+\alpha_\lambda)$.} (dashed line). 

Our best-fit slope is consistent with the one estimated by T02 in the FUV ($-0.69\pm0.06$ at $\lambda>1216$\AA, see their Table~1), and by S12 ($-0.68\pm0.14$) in the same region (see Figure~\ref{lum}). We note that the continuum regions adopted by S12 are the same as T02. 
\citet{vandenberk2001} also measured the continuum slope of a quasar composite utilising 2200 spectra from SDSS, which span a redshift range of $0.044\leq z \leq 4.789$. They find a slope of $\alpha_\nu=-0.44\pm0.10$ (no IGM absorption correction is applied), which is significantly different than the one we measured at 2.4$\sigma$. The difference can be attributed to the different continuum regions adopted by \citet{vandenberk2001} (i.e. 1350$-$1365 and 4200$-$4230\AA). 

We have also applied different continuum regions to gauge the
dependence of the fit on the windows adopted. We considered the
intervals listed by \citet{2010MNRAS.402.2441D} for a sample of 96
quasars at redshift lower than 3 with spectra collected from several
instrument (e.g., SDSS, ESO/NTT, Nordic Optical Telescope, and
HST). Specifically, the alternative windows adopted are 1351$-$1362,
1452$-$1480, 1680$-$1710, 1796$-$1834, 1970$-$2010, and
2188$-$2243\AA. To also fit the UV part of the spectrum, we add to
these the 5 intervals at $\lambda < 1200$\AA~by T02. We find a slope
of $-0.63 \pm 0.01$.

From Fig.\ref{fit} it is apparent that a single power-law provides an excellent fit up to $\sim900$\AA~ where the continuum exhibits a break. A flatter (softer) spectrum is present at shorter wavelength, which is not modelled by the simple single power-law.
Solely for the purposes of a quantitative estimate of the position of the break, we have modelled the stacked  spectrum by adding an exponential attenuation to the power-law as 
\begin{equation}
f_\lambda \propto \begin{cases} \lambda^{\alpha_\lambda}, & \text{if } \lambda\geq \lambda_b \\ 
\lambda^{\alpha_\lambda}\times e^{-(\lambda_b-\lambda)/\lambda_f}, & \text{if } 600\textrm{\AA}<\lambda<\lambda_b
\end{cases}
\end{equation}
where $\lambda_b$ and $\lambda_f$ represent the break and the attenuation wavelengths, respectively.
The best-fit power law index is $\alpha_\nu=-0.65\pm0.01$, the break occurs at $\lambda_b=922 \pm 31$ with an attenuation factor $\lambda_f\simeq450$ not well constrained given the large uncertainties on the stacked spectrum in the far blue.

The break measured from our stacked spectrum is at bluer wavelengths with respect to the ones estimated from composites at much lower redshift. S12 found the
break at $\lambda\sim1000$\AA, whilst T02 located the wavelength break
around 1200$-$1300\AA (a comparison among the various composites is given in Figure~\ref{lum}). 
The difference from the latter is due to the poor IGM correction at those wavelengths, which artificially produces a break around the \ion{Ly}{$\alpha$}. We will analyse this issue further in Section~\ref{Comparison with Previous Quasar Composites}.
Additionally, the composite shown by S04 is much steeper than what we observed in the same wavelength range (i.e. $\alpha_\nu=0.56^{+0.38}_{-0.28}$ at $\lambda=650-1150$\AA).  

\begin{figure}
 \centering\includegraphics[width=\linewidth,clip]{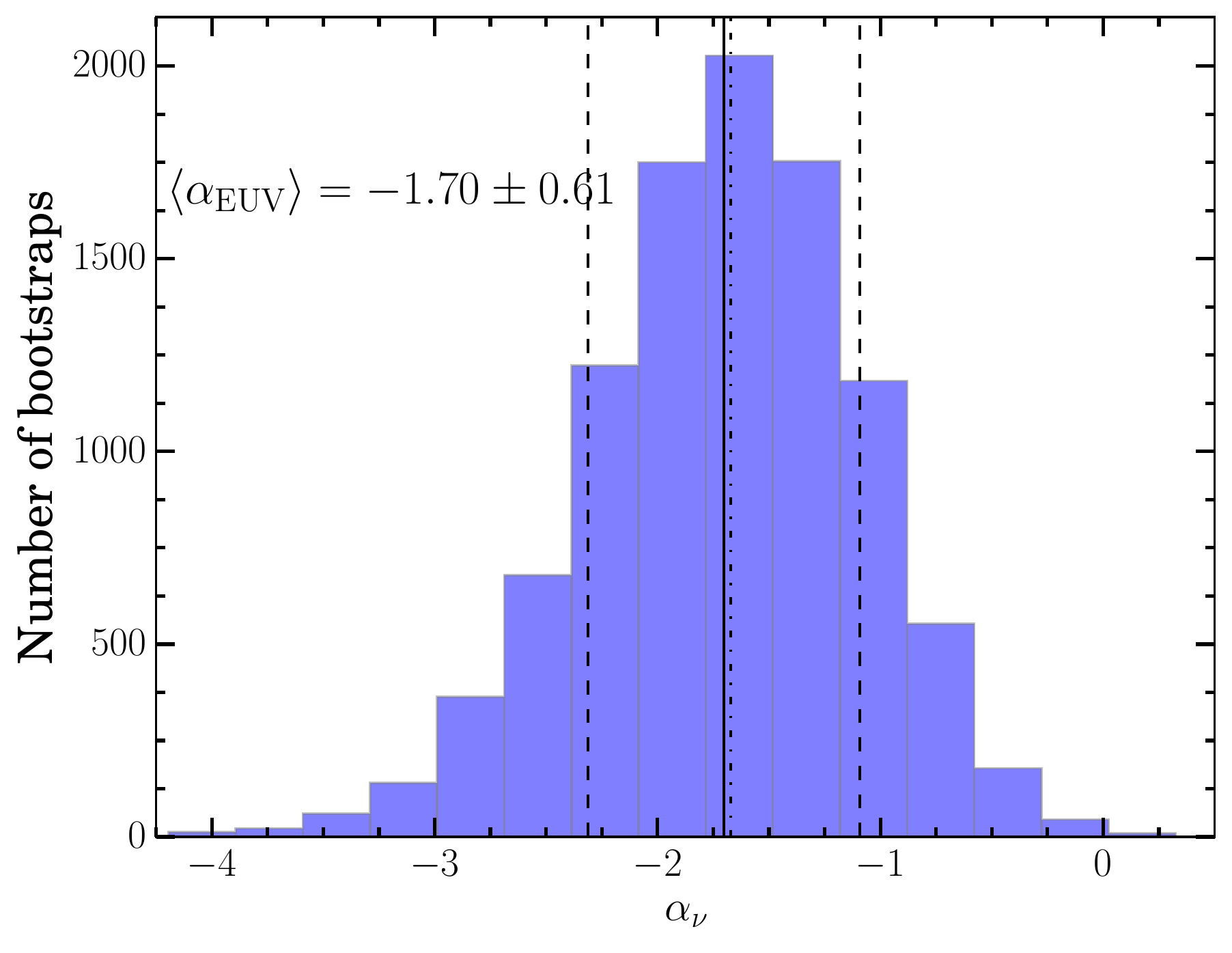}
 \caption{Histogram of the EUV slopes estimated from the procedure outlined in \S~\ref{Spectral fit and emission lines}. The solid and the dashed lines represent the mean and $1-\sigma$ dispersion, respectively, while the dot-dashed line denotes the median. }
 \label{hist_alphafuv}
\end{figure}

\begin{figure*}
\centering                                      
 \includegraphics[width=14cm]{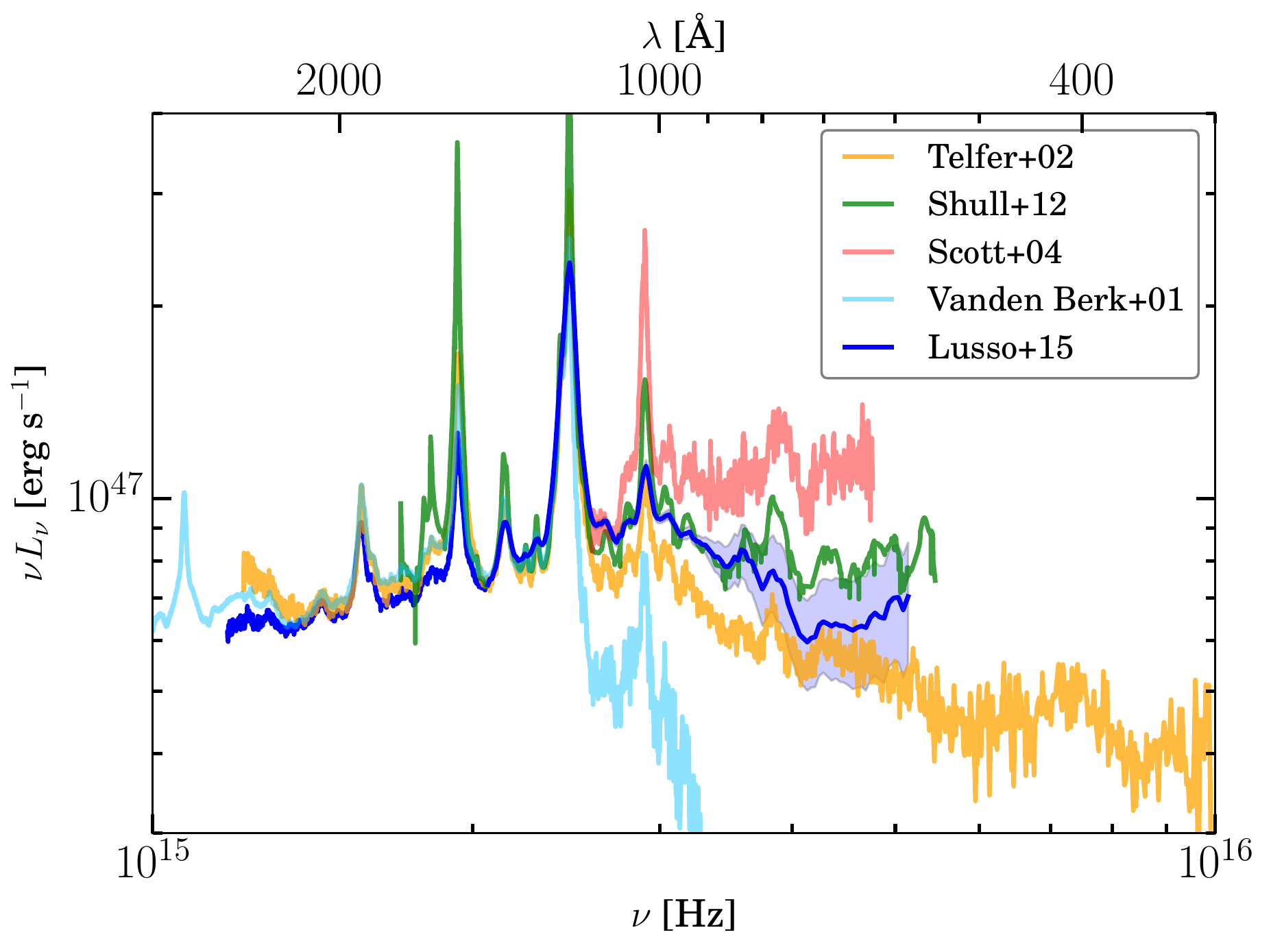}
 \caption{Mean IGM corrected QSO spectra with uncertainties (keys as in Fig.~\ref{wfc3sdss_stack} and \ref{igmfctcorr}) in the rest-frame $\log \nu-\nu L_\nu$ plane. Rest-frame wavelengths (in Angstrom) are plotted on the top x-axis. The T02, S12 and Vanden Berk et al. (2001) composites are normalized to our average spectrum at 1450\AA, and are shown for comparison with the orange, green, and cyan solid lines, respectively. The S04 quasar composite (light red solid line) is normalized to our stack at $\lambda=1114$\AA~($\log \nu=15.43$).}
 \label{lum}
\end{figure*}

 
\rev{Given that emission lines} are much fainter in the EUV region than in the FUV/NUV and that uncertainties far in the blue are significant, we have considered the full stacked spectrum (continuum + lines) at $\lambda<912$\AA\ and found that the EUV slope is $-1.65 \pm 0.07$. 
The error on the EUV spectral index from the fit of the average spectrum is quite small given the uncertainties on the WFC3 stack itself at these wavelengths. 
Additionally, the flux in the EUV systematically depends on the IGM correction applied, such that fluxes and errors are strongly correlated. The simple $\chi^2$ fitting technique is not appropriate in the EUV, and the bootstrap procedure is required.
We have thus estimated the EUV slope from each realization in the bootstrap described in Section~\ref{IGM transmission correction}. Figure~\ref{hist_alphafuv} shows the distribution of the EUV slopes of mock stacks. 
We found $\alpha_{\rm EUV} = -1.70 \pm 0.61$ (continuum + lines), which is a more reasonable value of the actual uncertainty. We will quote this as the final EUV slope.  \rev{Lastly, we comment that it is evident from the right panel of Fig.~\ref{fit} that a single power law does not seem to be a satisfactory description of the region below 912\AA. For example, there is no good reason why the feature at $\sim$ 730\AA\ should be considered an absorption feature. However, given the large uncertainties of our stack and to compare $\alpha_{\rm EUV}$ with previous evaluations from the literature, we refrain either to employ more complicated functions or to define continuum windows to fit this region.
We further discuss this issue in Sect.~\ref{Quasar photoionisation rate}.}

We also identify most of the emission lines in our stacked spectrum usually seen in optical spectra of high-redshift QSOs such as \ion{Ly}{$\beta$}, \ion{Ly}{$\alpha$}, \ion{Si}{iv}+\ion{O}{iv}], \ion{C}{iv}, and the semi-forbidden line of \ion{C}{iii}] 1909. 
A number of weak lines show up in the average spectrum at $\lambda<1216$\AA, including \ion{Ne}{Viii}+\ion{O}{iv} 772, \ion{O}{iii} 831, \ion{Ly}{$\gamma$}+\ion{C}{iii}] 873, and \ion{Fe}{ii}+\ion{Fe}{iii} 1123. Blended lines from high-ionisation states such as \ion{O}{iv} 608, \ion{O}{v} 630, \ion{N}{iii} 685, and \ion{O}{iii} 702, which are important diagnostics for investigating the physical conditions of broad emission line regions, may also be present, although it is impossible to reliably measure their strengths given the noise in our stacked spectrum at blue wavelengths. 
Weaker lines in the FUV include \ion{He}{ii} 1640, and \ion{O}{iii}] 1663. The properties of the emission lines for which the uncertainties on the stacked spectrum are small (i.e. $\lambda>912$\AA) are listed in Table~\ref{tab:emlines}. 

Summarising, we estimated the continuum slope in the NUV/FUV ($\alpha_{\nu} = -0.61 \pm 0.01$) and in the EUV ($\alpha_{\nu} = -1.70 \pm 0.61$).
We confirm the presence of a break in the quasar stacked spectrum by employing a completely independent sample at high redshift. We also emphasise that previous estimates of the break were performed with very few spectra ($\sim10$ in T02 and $3-12$ in S12) contributing at short wavelengths, whilst all 53 spectra in our high-redshift sample are contributing at 600\AA. 

\subsection{Comparison with Previous Quasar Composites}
\label{Comparison with Previous Quasar Composites}

In this section we will perform a detailed comparison between our $z_{\rm em}\sim2.4$ quasar average spectrum with previous works in literature. We caution the reader that a comprehensive and consistent explanation for all the differences hinges on multiple factors as, for example, sample selection biases. 

In Figure~\ref{lum} we compared our WFC3 average spectrum with the AGN composites found from SDSS by \citet{vandenberk2001}, 
from HST by T02, from COS by S12, and from FUSE by S04.  
The \citet{vandenberk2001}, T02, and S12 composites are normalized at 1450\AA, whilst the S04 composite is, instead, normalized to our IGM corrected mean stack at $\lambda=1114$\AA~($\log \nu=15.43$).
We caution that the latter normalization is more problematic, because it is subject to uncertainties in the \ion{Ly}{$\alpha$} forest IGM correction employed by S04.

Several important points emerge from the comparison in Figure~\ref{lum}. 
First, it is apparent that the IGM correction employed by T02 is significantly underestimated at $\lambda<1216$\AA. 
In general, T02 attempted to identify individual LLSs by eye and used estimates of the
Lyman limit optical depth to correct spectra with strong absorption. 
Whereas for lower column density systems, a single statistical 
correction for unidentified Lyman limit absorbers in the \ion{Ly}{$\alpha$} valley \citep{1990A&A...228..299M} was applied per
spectra using the column density distribution
\begin{equation}
\label{lyaabs}
\frac{\partial^2 n}{\partial z \partial N} \propto (1+z)^\gamma N^{-\beta},
\end{equation}
where $n$ is the number of lines, $z$ is the redshift, and $N$ is the column density of 
neutral hydrogen. Values for the $\beta$ factor considered by T02 are 
$\beta=1.83$ for $3\times10^{14}<\nh<10^{16}$ cm$^{-2}$ and $\beta=1.32$ for $\nh>10^{16}$ cm$^{-2}$ \citep{1993MNRAS.262..499P}.
For $3\times10^{12}<\nh<3\times10^{14}$ cm$^{-2}$ T02 used $\beta=1.46$ \citep{1995AJ....110.1526H}.

\begin{figure}
 \includegraphics[width=\linewidth,clip]{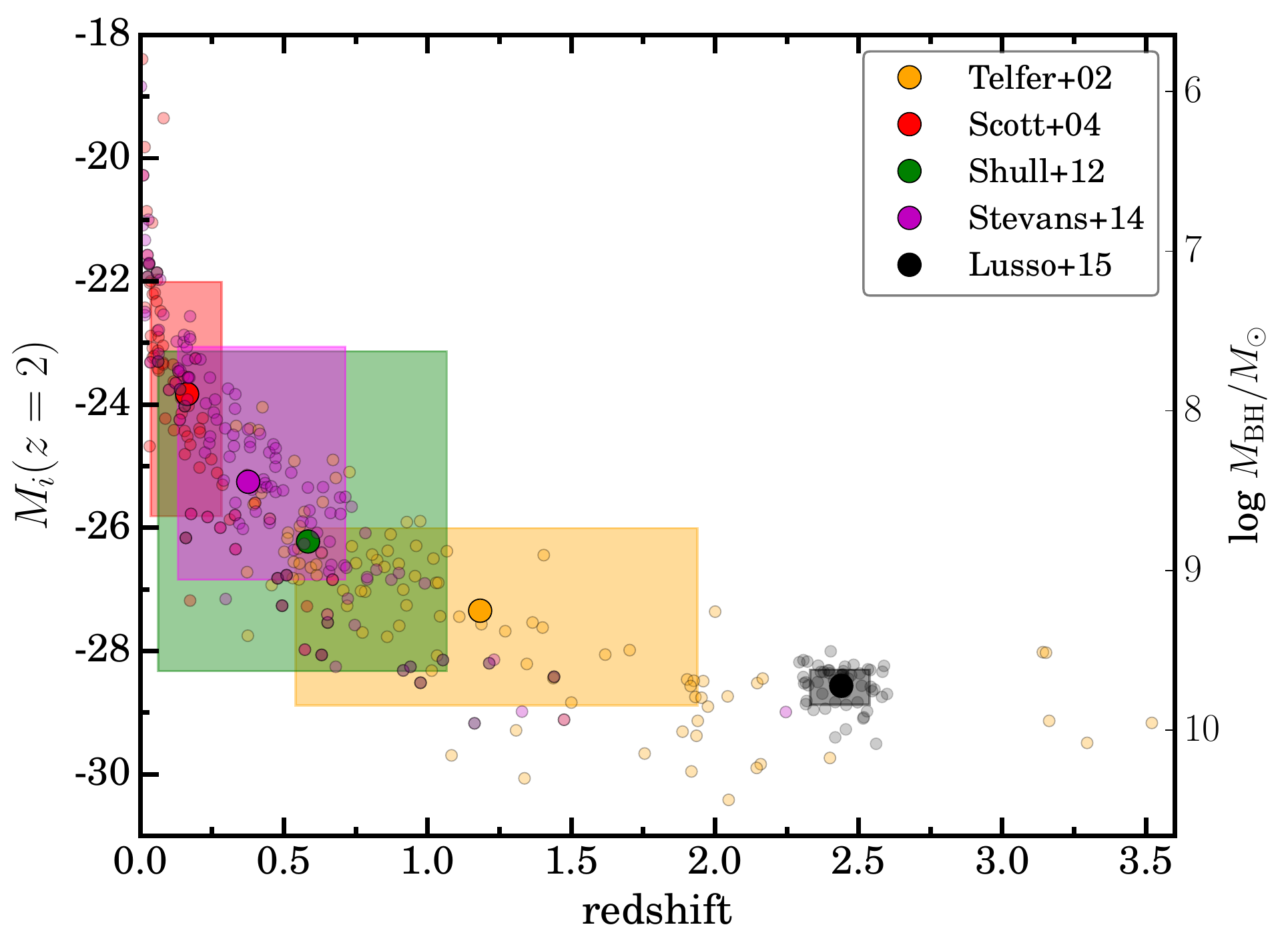}
 \caption{Absolute $i$-band magnitude (normalized at $z=2$, K-corrected following \citealt{2006AJ....131.2766R}) as a function of redshift. Shaded areas indicate the redshift and magnitude ranges for the different samples, estimated from the $16^{\rm th}$ and $84^{\rm th}$ percentiles. Large filled circles represent the median for the different samples: Our WFC3 sample (black), T02 (orange), S04 (red), S12 (green), and S14 (magenta). Indicative values of the black hole masses (in units of $M_\odot$) are plotted on the y-axis on the right. $\mbh$ is estimated via $\ledd = \lbol/\lumedd$ assuming the average $\ledd=0.35$ for the WFC3 sample.  We have estimated the relation between $\lbol$ and $\mi$ to be $\log \lbol = -10.03\mi/26 + 36.32$ by fitting the sources in the DR7 quasar catalog.}
 \label{magi_z}
\end{figure}

The application of this formula leads to a correction characterized by a stepwise behaviour, in which the opacity decreases toward shorter wavelengths (as already discussed in \citealt{2003ApJ...590...58B}). We computed the absolute $i$-band magnitude ($\mi$\footnote{The absolute $i$-band magnitudes normalized at $z=2$, K-corrected following \citealt{2006AJ....131.2766R}, for the WFC3 sample have been taken from the DR7 quasar catalog by Shen et al. (2011).}) for the T02 sample and these values are plotted as a function of redshift in Figure~\ref{magi_z}. There are about 20 quasars at $z > 2$ (contributing to the flux at $\lambda\sim300$\AA\ in the SED analysis) in the T02 sample and for these objects their correction is lower by a factor of $\sim 2$ in flux at $\lambda=600$\AA\ with respect to ours (see Fig. 3 in Binette et al. 2003).


The spectral slope measured by S04 in the EUV is $\alpha_\nu = -0.56\pm^{0.38}_{0.28}$, which is at variance with the spectral shape shown by our WFC3 stacked spectrum ($\alpha\simeq-1.7$ and with the other spectra from different samples) in the same wavelength range. Like T02, S04 adopted a similar procedure to correct for IGM absorption. LLSs were individually identified by eye and a single statistical correction for the line-of-sight absorption due to the Ly$\alpha$ forest and the Lyman valley was applied on single spectra as in Eq.(\ref{lyaabs}). At variance with T02, the parameters to correct for the Ly$\alpha$ forest absorption are $\beta=2$ for $1.6\times10^{12}<\nh<2.5\times10^{14}$ cm$^{-2}$, and $\beta=1.35$ for $2.5\times10^{14}<\nh<5\times10^{16}$ cm$^{-2}$. For the redshift distribution parameter S04 used $\gamma=0.15$.
This different parametrization leads to an IGM correction that goes from 5\% to 10\% in the wavelength range 1200$-$600\AA~ at $z\sim0.16$ (see their Fig.~4).  
Any judgment about this IGM correction being underestimated is not trivial given the lower redshifts of the S04 AGN sample, compared with those of the WFC3 sample as shown in Figure~\ref{magi_z}. However, biases in the S04 sample, and subtleties in correcting for LLSs may contribute in the differences between the FUSE and the WFC3 spectras.

We have also compared our WFC3 average spectrum with the one by S12. 
The COS composite does not significantly differ from our spectrum within the uncertainties, but this similarity cannot be easily explained given the different average redshift, the sample selection\footnote{The COS sample was selected from the archive as the best available high-S/N spectra of AGNs in January 2011.}, and the IGM correction employed by S12, which is solely performed by visual spectral inspection.
The advantage of the S12 sample is the high S/N and the high spectral resolution, which allow them to fit the local continua of their objects (correcting for identify LLSs). 
Although S12 missed some partial LLSs with small Lyman continuum optical depths, the composite constructed by S14 does not significantly differ from the one in S12. This is due to the fact that the partial LLSs that S14 found have very low columns, $\log \nh < 16$ (mostly in the 15.0-15.5 range) and their Lyman continuum opacity was negligible.  Additionally, S14 used the pattern of higher-Lyman series lines rather than the Lyman edge. The fitted EUV composites of S12 and S14 are in fact quite similar: $\langle \alpha_\nu \rangle  =  -1.41 \pm 0.21$ in S12, while S14 found $-1.41 \pm 0.15$.
We also note that, in the S14 analysis only $\sim20$ spectra contribute at 700\AA, while few of them do not cover 912\AA\ such that the continuum normalization may be problematic.  
Given the bias in the UV samples, and the different approach in the IGM absorption correction, it is unclear what significance to attach to this agreement.

\begin{table}
  \caption{EW comparison}
  \label{tab:baskin}
  \begin{center}
    \leavevmode
    \begin{tabular}{lccccc} \hline \hline              
    Line  & ionisation         & \multicolumn{4}{c}{EW(line) / EW(\ion{Ly}{$\alpha$}) } \\   
          & energy (eV)        & \multicolumn{4}{c}{$\alpha_{\rm ion}$}   \\
            &      &  $-1.2$  & $-1.6$ & $-2.0$ & $-1.7$ (Lusso+15) \\ \hline      
\ion{Ly}{$\alpha$} & 13.6 &  1.  & 1.  & 1.   & 1.  \\
  \ion{C}{iii}]      & 24.0 &  0.1  & 0.10 & 0.13  & 0.21  \\
  \ion{He}{ii}       & 54.4 & 0.09  & 0.06 & 0.03  & 0.03  \\
  \ion{C}{iv}        & 48.0 & 0.78  & 0.44 & 0.25  & 0.24  \\
     \hline
    \end{tabular}
  \end{center}
\end{table}
\begin{figure}                                   
 \includegraphics[width=\linewidth]{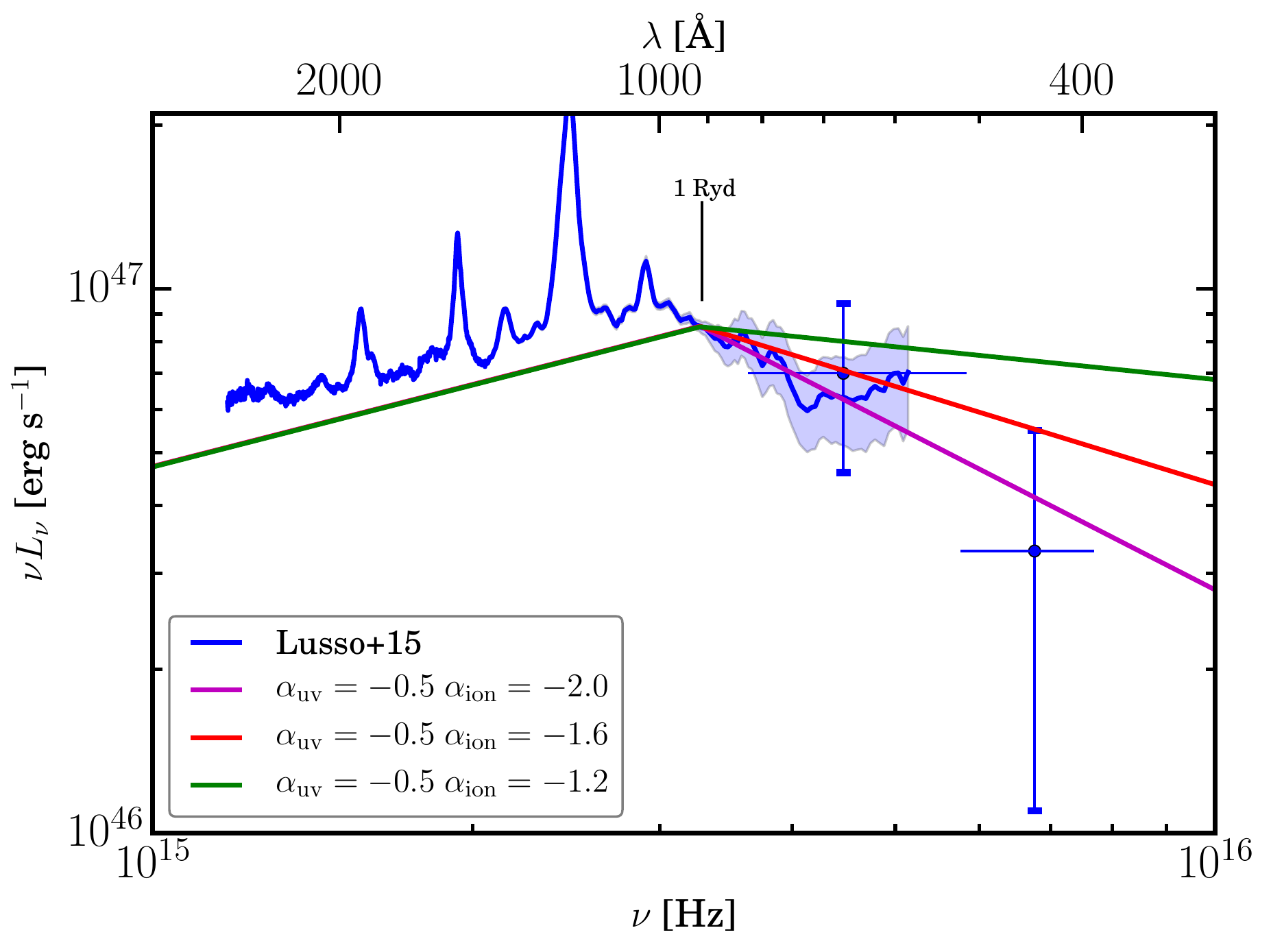}
 \caption{Comparison between the three types of SEDs adopted by \citet{2014MNRAS.438..604B} with our WFC3 average spectrum. The green, red, and magenta solid lines correspond to $\alpha_{\rm ion} = -1.2$, -1.6 and -2.0, respectively. The blue points represent the IGM corrected mean GALEX forced photometry, where horizontal bars indicate the GALEX band-passes.}
 \label{photo}
\end{figure}

\subsection{Comparison with photoionisation models}
\label{Comparison photoionisation models}

In this Section we compare the line ratios observed in the WFC3 average spectrum and the line ratios predicted by photoionisation models. 
\citet{2014MNRAS.438..604B} presented a radiation-pressure-dominated hydrostatic solution for gas in the broad line region, using the photoionisation code CLOUDY \citep{1998PASP..110..761F}
We defer the reader to that paper for details of the physical model. Here, we compare their predicted line ratios with our estimates, for several prominent BLR emission lines. 
The predicted line ratios are based on three types of SEDs, differing from each other in the ionising slope $\alpha_{\rm ion}$ at energies between 1 Rydberg and 1 keV (see their Section 2.2). 
These three SEDs are plotted in Figure~\ref{photo}. We compare the observed and predicted line ratios, rather than the equivalent widths (EW) of single lines, since the average covering factor of the BLR in the WFC3 sample could be different than the covering factor of 0.3 assumed by Baskin et al. In order to calculate the predicted line ratios, we use the Baskin et al. model with solar metallicity, and a BLR distance where the line emissivity peaks (see their Figure 5), which is roughly the expected EW if one assumes a BLR which spans a range of distances. 
The predicted \ion{C}{iii}]/\ion{Ly}{$\alpha$}, \ion{He}{ii}/\ion{Ly}{$\alpha$}, and \ion{C}{iv}/\ion{Ly}{$\alpha$} for $\alpha_{\rm ion}=-1.2$, $-$1.6 and $-$2.0 are listed in Table~\ref{tab:baskin}, together with the observed line ratios. 
The observed \ion{He}{ii}/\ion{Ly}{$\alpha$} and \ion{C}{iv}/\ion{Ly}{$\alpha$} are consistent with the $\alpha_{\rm ion}=-2.0$ model, which is softer than our derived slope of -1.7, though within the uncertainties. 
The observed \ion{C}{iii}]/\ion{Ly}{$\alpha$} is a factor of two larger than expected for all assumed values of $\alpha_{\rm ion}$.

\subsection{Comparison with GALEX}
\label{Comparison with GALEX}

We have also increased our coverage at shorter wavelengths by considering the GALEX photometry of the WFC3 sample (see Section~\ref{GALEX}). These data have been corrected from IGM absorption as follows
\begin{enumerate}
 \item We generate a set of 6000 mock quasar samples, following a similar
 procedure as in Section~\ref{Composite construction}. Each sample contains 53 NUV and FUV fluxes (we again allow for duplications), and we compute the mean of these NUV and FUV fluxes.
 \item We normalize the average GALEX fluxes to rest frame 1450\AA\ (as for the WFC3 spectra). \rev{The 1450\AA\ flux is estimated from each average WFC3 spectrum of the mock quasar samples}.
 \item We then randomly draw one IGM transmission function from our suite of 10,000 and we integrate this function over the GALEX filter curves. This is repeated for each mock quasar sample.
 \item The 6000 mock stacks are thus corrected from IGM absorption and averaged to produce the final mean NUV and FUV fluxes.
 \item The GALEX fluxes are then multiply by the average flux at 1450\AA\ of the WFC3 sample.
 \item The uncertainties on the corrected GALEX fluxes are estimated from the dispersion of these 6000 mock stacks.
\end{enumerate}
Both IGM absorption corrected NUV and FUV bands are plotted in Figure~\ref{photo}. 
We ignore quasar variability when stacking and plotting the results together with the stacked spectra.
The GALEX NUV band almost covers the WFC3 spectra in the Lyman continuum, hence the fact that the NUV flux is consistent with the WFC3 stack within the uncertainties further check our results.

Nonetheless, the error bars on the NUV and FUV fluxes are significant and do not allow us to further constrain any of the three SEDs, even though the FUV flux seems to exclude the SED with $\alpha_{\rm ion}=-1.2$ consistently with the comparison with the predicted EWs.

\begin{figure*}
\centering
\includegraphics[width=14cm]{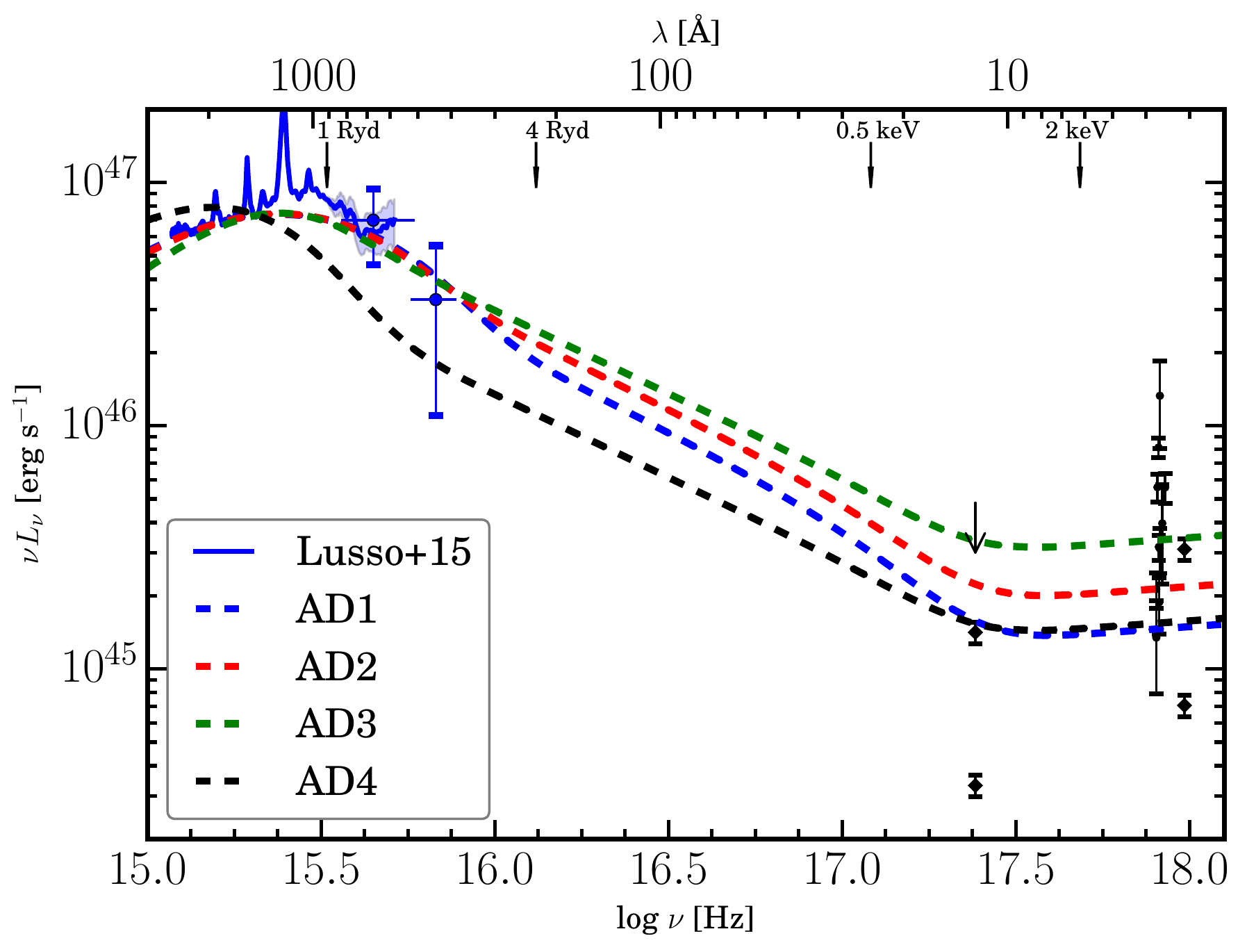}
 \caption{Accretion disc models. Keys as in Figure~\ref{photo}. The blue (AD1), red (AD2), and green (AD3) dashed lines represent the three reference accretion disc models for $\mbh=6\times10^9$ with $\ledd=0.35$ ($a=0.8$), $3\times10^9M_\odot$ with $\ledd=0.70$ ($a=0.3$), and $1.2\times10^{10}$ with $\ledd=0.17$ ($a=1.0$), respectively. The black dashed line is the same as AD1 but with $a=0$ and $r_{\rm corona}=20$ (AD4). All models are normalised to 1450\AA. X-ray data for the 10 quasars detected by ROSAT are plotted with filled black points. The XMM-Newton soft and hard luminosities for J0755$+$2204 and  for J1119$+$1302 are plotted as black diamonds. The black arrow represent the ROSAT flux limit at 1 keV.}
 \label{disc}
\end{figure*}
\begin{table}
  \caption{Accretion disc model parameters}
  \label{tab:disc}
  \begin{center}
    \leavevmode
    \begin{tabular}{lcccc} \hline \hline              
    Ref. model  & $\mbh$   & $\ledd$  & $r_{\rm corona}$ & $a$\\   
                       & ($M_\odot$) &         &  ($R_{\rm g}$)      &     \\ \hline
  AD1  (blue)          & $6\times10^9$        &  0.35  & 8 & 0.8    \\
  AD2  (red)       & $3\times10^9$        &   0.70  &  20 & 0.3    \\
  AD3  (green)           & $1.2\times10^{10}$ &  0.17  &  8 &  1.0    \\
  AD4  (black)        & $6\times10^9$        &   0.35  & 20 & 0.0   \\
   \hline
    \end{tabular}
 \flushleft\begin{list}{}
 \item {\bf Notes.}
 \item { All models have the following parameters fixed: $\Gamma=1.9$, $k T_{\rm e}=0.15$ keV, $f_{\rm pl}=0.2$, $\tau=18$, and $r_{\rm out}=10^5 R_{\rm g}$.}
 \end{list}
  \end{center}
\end{table}
\subsection{AGN Accretion disc models}
\label{Accretion disc models}

A basic prediction of simple accretion disc models is that the disc temperature decreases as the black hole mass increases \citep{1973A&A....24..337S}, thus
\begin{equation}
\label{tdisc}
T = \left(\frac{G M \dot{M}}{4 \pi \sigma r^3} \right)^{1/4} \sim \\
6.3\times10^5 \left(\frac{\dot{M}}{\dot M_E}\right)^{1/4} M_8^{-1/4} \left(\frac{r}{2 R_g}\right)^{-3/4},
\end{equation}
where $\dot{M}/{\dot M_E}$ is the Eddington ratio ($\ledd$, the accretion rate normalized to the Eddington accretion rate), $M_8=\mbh/10^8M_\odot$, and $R_g$ is the gravitational radius ($R_g = G M/c^2$).
For a scale $r$ of $6R_g$ and an Eddington ratio $\ledd=0.1$, the disc temperature goes from $\sim5\times10^5$K to $\sim8.7\times10^4$K for a $\mbh$ of $10^6$ and $10^9M_\odot$, respectively\footnote{Equation~(\ref{tdisc}) is valid in the Newtonian limit, and therefore may be not accurate to estimate the temperature in the disc, which is highly relativistic. See Eqs. (3)-(5) and Table 1 in \citet{2011MNRAS.417..681L} for a more accurate approach.}.
The disc temperature sets the peak of the BBB and, thus, we expect to see the location of the break changing as a function of $\mbh$.

The UV composites by T02 and S12 show a break between \ion{Ly}{$\alpha$} and 1000 \AA, whilst the quasar composite by S04 does not show any. This might be consistent with the fact that the S04 is the lowest luminosity/redshift sample and presumably it has a lower $\mbh$ on average than the S12 and T02 samples. 
Unfortunately, black hole mass values are not available for most sources in the T02, S04 and S12 samples. S04 have compiled black hole mass values from the literature for 22 objects in their sample (with a median $\mbh$ of $1.6\times10^8M_\odot$) and they found a significant correlation between the spectral index and black hole mass, with the spectral slope being softer for higher $\mbh$, which may run in the expected direction. However, as we have discussed in the previous Sections, both the presence and the position of the break depend strongly on the IGM correction considered.

Furthermore, the standard black body disc model depicted above, does not reproduce the observed soft X--ray emission seen in AGN, consequently other physical parameters are required such as black hole spin and/or substantial Compton upscattering (\citealt{1997ApJ...477...93L}, but see also \citealt{2015MNRAS.446.3427C}). 

We further analysed our WFC3 average spectrum in the context of accretion disc models. In particular we have considered the publicly available energetically self-consistent model, \textsc{optxagnf}, developed by \citet{2012MNRAS.420.1848D} within the \textsc{XSPEC} spectral fitting package.
Their model contains three distinct spectral components, all powered by the energy released by a single accretion flow of constant mass accretion rate, $\dot M$, onto $\mbh$: an outer and inner disc, and an X--ray corona. The outer disc emits as a (colour temperature corrected) blackbody, the inner disc is where a fraction of the disc emission is Compton upscattered, while the X--ray corona is where also a fraction of the emission is Compton upscattered producing the power-law tail at high energies (see Section 4 in \citealt{2012MNRAS.420.1848D} for further details). 
This model is set by 9 parameters: $\mbh$, $\ledd$, black hole spin ($a$), radius of the X--ray corona ($r_{\rm corona}$), outer accretion disc radius ($r_{\rm out}$), electron temperature ($k T_{\rm e}$), optical depth ($\tau$) of the Comptonization region, photon index ($\Gamma$), and the fraction of energy dissipated as a hard X--ray  power law from $r_{\rm corona}$ to the innermost stable circular orbit $r_{\rm isco}$ ($f_{\rm pl}$).

Our data deliver constraints on $\mbh$ and $\ledd$, which have been
collected from the DR7 quasar catalog (\citealt{2011ApJS..194...45S})
for all quasars in the WFC3 sample. The average $\mbh$ estimated from
the \ion{C}{iv} line is $6\times10^9M_\odot$, whilst the Eddington
ratio is 0.35. The reliability of utilising the \ion{C}{iv} line is
controversial, since \ion{C}{iv} can be severely affected by
non-virial motions such as outflows, winds, and strong absorption
(e.g. \citealt{2005MNRAS.356.1029B,2008ApJ...680..169S}). We will refer to this model as
AD1. We have thus investigated two additional models with half (AD2)
and double (AD3) the average $\mbh$, where the average $\ledd$ has been
re-scaled consequently. 
The photon index has been fixed to 1.9, which
is usually observed in quasars (e.g. \citealt{2005A&A...432...15P}).  For the
other parameters we do not have any constraint from the data, and
hence we have to assume values in order to reproduce both NUV/FUV/EUV and
X--ray in a reasonable way. We have thus fixed the electron
temperature $k T_{\rm e}$ at 0.15 keV \citep{2004MNRAS.349L...7G},
$f_{\rm pl}$= 0.2, $\tau=$18, and the default outer accretion disc
radius, $r_{\rm out}=10^5 R_{\rm g}$. The parameter values for the
reference models are given in Table~\ref{tab:disc}, and plotted in
Figure~\ref{disc}. All models are normalised to 1450\AA. 
We note that all the models are only meant to provide a reasonable description
of the observables, since we did not attempt any simultaneous fit of the 
optical-UV/X--ray data.
X-ray data for the ten quasars detected by ROSAT and
XMM-Newton are shown in Figure~\ref{disc} for completeness. 

A key aspect of these models is that the peak of the BBB is set by the black hole mass, spin and mass accretion rate through the outer disc. 
Interestingly, the AD4 model (i.e. average $\mbh$ and $\ledd$ and spin zero) do not provide a good representation of the FUV/EUV region of our average spectrum\footnote{The radius of the corona has been fixed to 20 $R_g$, instead of 8 as in AD1.}, with the BBB being colder than observed (see also \citealt{2012MNRAS.423..451L}). 
This problem can be solved if we instead consider the AD2 but with the BH spin set to zero. 
If the BH spin increases, the distance of the innermost stable circular orbit reduces, and this effects the disc temperature as described by Eq.\ref{tdisc}.

Summarising, if we consider the BH masses from \ion{C}{iv} to be correct, we need to have non zero values of the BH spin to match the data, with higher values of the BH spin as the $\mbh$ increases (see \citealt{2014ApJ...789L...9T} for similar results, i.e. black hole masses above $\sim 3\times10^9 M_\odot$ at $z\sim2-3$ are consistent with very high spins).
We do not need BH spin if we instead consider a factor of 2 mass lower, on average.
The latter is in agreement with previous results that have found BH mass values from \ion{C}{iv} systematically higher, once calibrated with virial mass estimators based on the \ion{Mg}{ii} (e.g. \citealt{2008ApJ...680..169S,2013ApJ...770...87P}, but see also \citealt{2012MNRAS.427.3081T}). 

In principle, the combination of reliable $\mbh$ and FUV/EUV spectra (properly corrected for IGM absorption) may provide hints on the quasar BH spin. 

It is not possible to test if the location of the break changes as a function of $\mbh$ within the WFC3 since it has a narrow luminosity range. 
We thus need to have a comparison sample at low redshift/luminosity, hence lower $\mbh$. The FUSE composite by S04 is constructed with low redshifts/luminosities AGN (see Fig.\ref{magi_z}) and interestingly it shows a hard spectral slope in the EUV with no break consistent with the BBB being produced by small $\mbh$. This might point towards the direction of the predicted trend, but further studies are needed. 
In particular, a natural extension of the present analysis is to cover bluer wavelengths with high resolution quasar spectra.

\subsection{Quasar emissivity and photoionisation rate} 
\label{Quasar photoionisation rate}

The UV background is governed by the ionising emissivities of the relevant source populations (typically quasars and star-forming galaxies) and cosmological radiative transfer in the IGM \citep[e.g.][]{faucher09,haardt12}. The specific comoving emissivity of the quasar population at redshift $z$ and frequency $\nu$,
\begin{equation}\label{defemissivity}
\epsilon_\nu\left(\nu,z\right)=\int_{L_\mathrm{min}}^{\infty}\phi\left(L,z\right)L_\nu\left(L,\nu\right)\mathrm{d}L\quad,
\end{equation}
critically depends on the adopted quasar luminosity function $\phi\left(L,z\right)$, its faint end ($L_\mathrm{min}$), and the quasar SED ($L_\nu\left(L,\nu\right)$). In particular, since the quasar luminosity function is typically determined in the NUV-optical, the specific emissivity at the Lyman limit $\epsilon_{\nu,912}$ depends on the chosen FUV and EUV SED parameterization, which is commonly taken as a luminosity-independent broken power law based on quasar composite spectra (Z97, T02). Resulting uncertainties in the specific emissivity are rarely discussed (see \citealt{faucher08} for an exception), or limited to variations in the EUV spectral index \citep{faucher09}. Consequently, current estimates of the specific Lyman limit emissivity of quasars still vary by a factor $\sim 2$ at $z=3$--4 \citep{hopkins07,siana08,cowie09,masters12}.

\begin{figure}
 \includegraphics[width=\linewidth,clip]{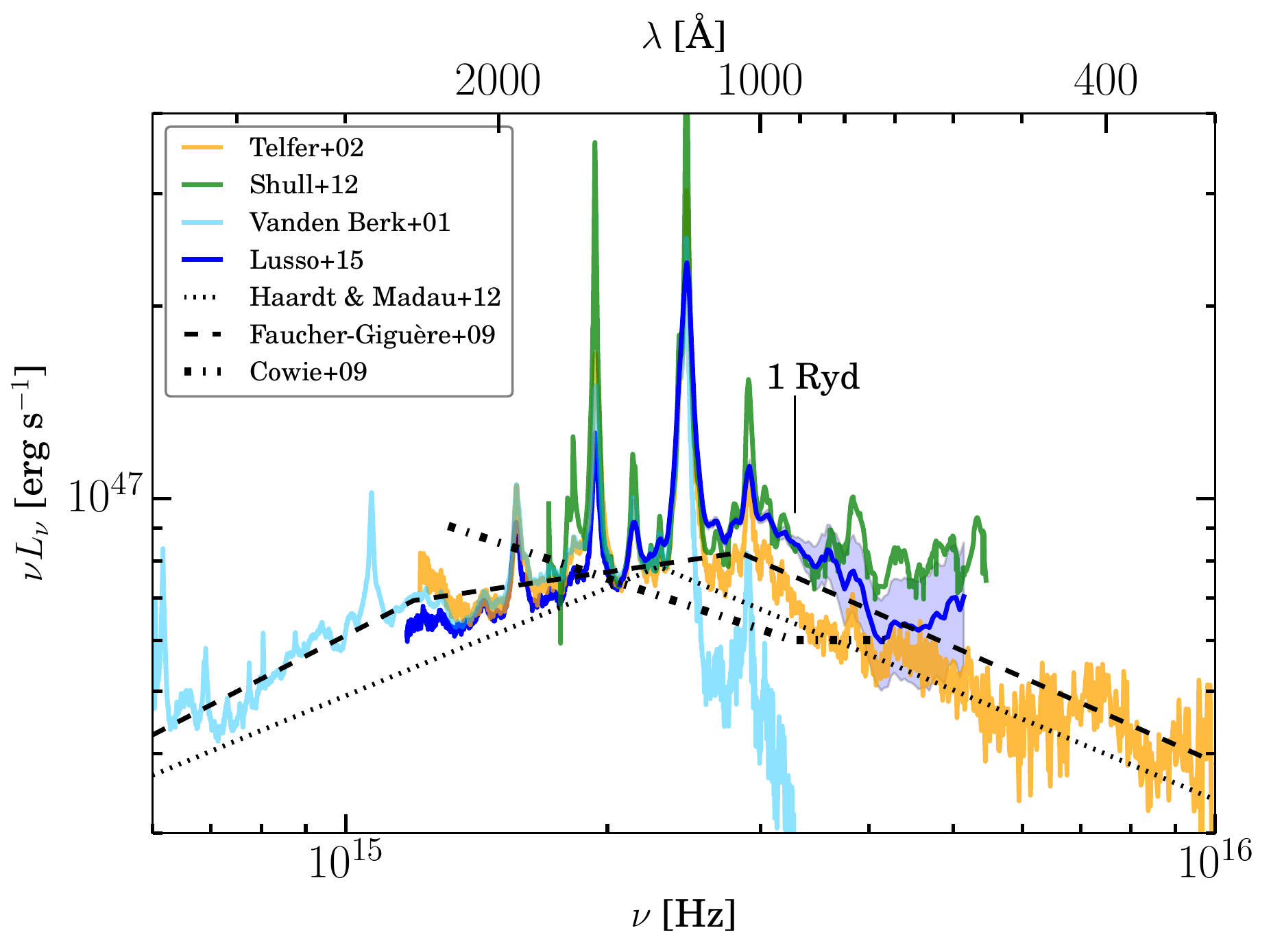}
 \caption{ As Fig.~\ref{lum} but including recent broken power law parameterizations of the quasar continuum \citep{cowie09,faucher09,haardt12}. The SEDs have been normalized at the adopted rest frame wavelengths (\citealt{faucher09}: 4400\,\AA) or assumed at 1450\,\AA\ if not given by the authors \citep{cowie09,haardt12}.
All parameterizations underestimate the quasar Lyman limit flux of our stacked spectrum and recent quasar composite spectra (S12, S14).}
 \label{sed_comp}
\end{figure}

Figure~\ref{sed_comp} compares our stacked spectrum and various quasar composites to broken power-law parameterizations used to estimate the specific quasar Lyman limit emissivity and/or the quasar contribution to the UV background \citep{cowie09,faucher09,haardt12}. Most previous SED parameterizations have underestimated the quasar Lyman continuum flux because they have been based on the T02 composite spectrum without a proper IGM correction (Section~\ref{Comparison with Previous Quasar Composites}). Adopting our power-law fit $f_\nu\propto\nu^{-0.61}$ instead, we obtain Lyman limit flux ratios of $f_{\nu,912}^\mathrm{L14}/f_{\nu,912}^\mathrm{FG09}=1.12$ \citep{faucher09} and $f_{\nu,912}^\mathrm{L14}/f_{\nu,912}^\mathrm{HM12}=1.38$ \citep{haardt12} as correction factors to the quasar Lyman limit emissivity $\epsilon_{\nu,912}(z)$ for a fixed luminosity function\footnote{Note that \citet{haardt12} simplistically assume $\epsilon_{\nu,912}(z)$ and $L_\nu(\nu)$ to be independent.}. 
\rev{We point out that \citet{cowie09} computed a power-law slope of --1.35 at $700<\lambda<2300$\,\AA\ obtained from GALEX photometry of a $z_\mathrm{em}\sim 1$ AGN sample. This implies a 700\AA\ to 2300\AA\ flux ratio of 0.20. Cowie et al. then assumed the actual SED below the break to be $f_\nu\propto\nu^{-1}$, which returns an extrapolated flux ratio $f_{\nu,912}/f_{\nu,2300}=0.26$. In Fig.~\ref{sed_comp} we have taken the break at exactly 912\AA\footnote{\rev{The broken power-law quasar SED is then $f_\nu\propto\nu^{-1.45}$ ($f_\nu\propto\nu^{-1}$) at $\lambda>912$\,\AA\ ($\lambda<912$\,\AA)}}. 
We also note here that the mean 2300\AA\ luminosity for the objects in the Cowie et al. sample is much lower ($\nu L_\nu \simeq 5\times10^{44}$ erg s$^{-1}$) than the one of our WFC3 sample.
}
Adopting a standard concave SED shape instead, their $\epsilon_{\nu,912}$ values increase by a factor of $\simeq 2$. While this correction makes their values consistent with higher estimates of $\epsilon_{\nu,912}$ \citep{hopkins07,siana08,masters12}, we note that \citet{siana08} and \citet{masters12} assume a low flux ratio $f_{\nu,912}/f_{\nu,1450}=0.58$ based on the T02 composite to convert their $\epsilon_{\nu,1450}$ to $\epsilon_{\nu,912}$. Our SED yields a 30\% increase to their estimates, such that $\epsilon_{\nu,912}\simeq 5\times 10^{24}$\,erg\,s$^{-1}$\,Hz$^{-1}$\,Mpc$^{-3}$ at $z\sim 3.2$. With this correction, quasars may account for the total specific Lyman limit emissivity estimated from the Ly$\alpha$ forest at $z\la 3.2$ \citep{becker13}, although a comparable contribution from star-forming galaxies is allowed within the current large uncertainties \citep{nestor13,mostardi13,becker13}. The quasar contribution seems to decrease substantially to $z\sim 4$, as the low (corrected) value of $\epsilon_{\nu,912}\simeq 1.8\times 10^{24}$\,erg\,s$^{-1}$\,Hz$^{-1}$\,Mpc$^{-3}$ from \citet{masters12} is just $\sim 20$\% of the total value inferred from the Ly$\alpha$ forest (\citealt{becker13}, although see \citealt{glikman11} for a $\sim 15$ times higher estimate of $\epsilon_{\nu,1450}$ at $z\sim 4$).

From the Lyman limit emissivity we can estimate the quasar contribution to the UV background photoionisation rate
\begin{equation}\label{defgamma}
\Gamma_\textrm{\ion{H}{i}}\left(z\right) = \int^{\infty}_{\nu_{912}}\frac{4 \pi J_\nu\left(\nu,z\right)}{h\nu}\sigma_\mathrm{HI}\left(\nu\right)\mathrm{d}\nu\quad,
\end{equation}
where $J_\nu\left(\nu,z\right)$ is the mean intensity of the ionising background at redshift $z$ and $\sigma_\mathrm{HI}\left(\nu\right)\simeq\sigma_{912}\left(\nu/\nu_\mathrm{912}\right)^{-3}$ is the \ion{H}{i} photoionisation cross-section with $\sigma_{912}=6.33\times10^{-18}$\,cm$^2$ at the Lyman limit frequency $\nu_{912}$.
At high redshifts $z\ga 4$ the mean free path of Lyman continuum photons in the IGM, $\lambda_\mathrm{mfp}\left(\nu,z\right)$, is much shorter than the Horizon, such that the UV radiation field at any given point is dominated by local sources within a sphere of radius $\lambda_\mathrm{mfp}$, and cosmological redshift effects in $J_\nu\left(\nu,z\right)$ can be neglected. In this `local-source approximation' the intensity of the UV background is
\begin{equation}\label{defuvb}
J_\nu\left(\nu,z\right)\simeq\frac{1}{4\pi}\lambda_\mathrm{mfp}\left(\nu,z\right)\left(1+z\right)^3\epsilon_\nu\left(\nu,z\right)
\end{equation}
\citep{madau99,schirber03,faucher08,becker13}, with $\lambda_\mathrm{mfp}$ ($\epsilon_\nu$) given in proper (comoving) units.
The proper mean free path to Lyman limit photons is well described by a power law
\begin{equation}\label{defmfp}
\lambda_\mathrm{mfp,912}=37\left(\frac{1+z}{5}\right)^{-5.4}\,\mathrm{Mpc}
\end{equation}
 at $2.3<z<5.5$ \citep{2014MNRAS.445.1745W}, whereas its frequency dependence is
\begin{equation}\label{defmfp2}
\lambda_\mathrm{mfp}\left(\nu,z\right)=\lambda_\mathrm{mfp,912}\left(\frac{\nu}{\nu_{912}}\right)^{1.5}
\end{equation}
 for a column density distribution $\fnz\propto\nh^{-1.5}$ \citep{faucher08}\footnote{Note that this is an approximation due to departures of $\fnz$ from a single power law \citep[e.g.][]{prochaska10,haardt12}.}. With our power-law parameterization of the quasar SED\footnote{We stress that due to the low resolution we cannot fit the EUV continuum beneath existing EUV emission lines (S12, S14). For computing the photoionisation rate, this turns into an advantage, as we approximately account for the EUV lines that also contribute ionising photons. With sufficient wavelength coverage and spectral resolution, the SED should be integrated instead of the power-law continuum.} $f_\nu\propto\nu^{\alpha_\mathrm{EUV}}$ at $\lambda>912$\,\AA, we can write the comoving specific emissivity of quasars as
\begin{equation}\label{defemissiv}
\epsilon_\nu\left(\nu,z\right)=\epsilon_{\nu,912}\left(z\right)\left(\frac{\nu}{\nu_{912}}\right)^{\alpha_\mathrm{EUV}}.
\end{equation}
Combining Equations~\eqref{defuvb}--\eqref{defemissiv} and integrating Equation~\eqref{defgamma}, the quasar contribution to the UV background photoionisation rate is
\begin{eqnarray}\label{defgammaq}\nonumber
\Gamma_\textrm{\ion{H}{i}} &\simeq& 4.6\times 10^{-13}\,\mathrm{s}^{-1}\left(\frac{\epsilon_{\nu,912}}{10^{24}\,\mathrm{erg}\,\mathrm{s}^{-1}\,\mathrm{Hz}^{-1}\,\mathrm{Mpc}^{-3}}\right)\\
& &\times\left(\frac{1+z}{5}\right)^{-2.4}\frac{1}{1.5-\alpha_\mathrm{EUV}}.
\end{eqnarray}
As pointed out by \citet{becker13}, the local-source approximation becomes increasingly inaccurate towards lower redshifts due to redshifting of Lyman limit photons. Consequently, Equation~\eqref{defgammaq} overestimates $\Gamma_\textrm{\ion{H}{i}}$ for a given $\epsilon_{\nu,912}$. This may be significant already at $z\sim 4$, since the quasar \ion{H}{i} photoionisation rate estimated from Equation~\eqref{defgammaq} with our $\alpha_\mathrm{EUV}=-1.7$ and the rescaled emissivity from \citet{masters12} is somewhat higher than the value obtained by \citet{haardt12} with approximately the same emissivity and mean free path but including redshift and radiative transfer effects.

\begin{figure}
 \includegraphics[width=\linewidth,clip]{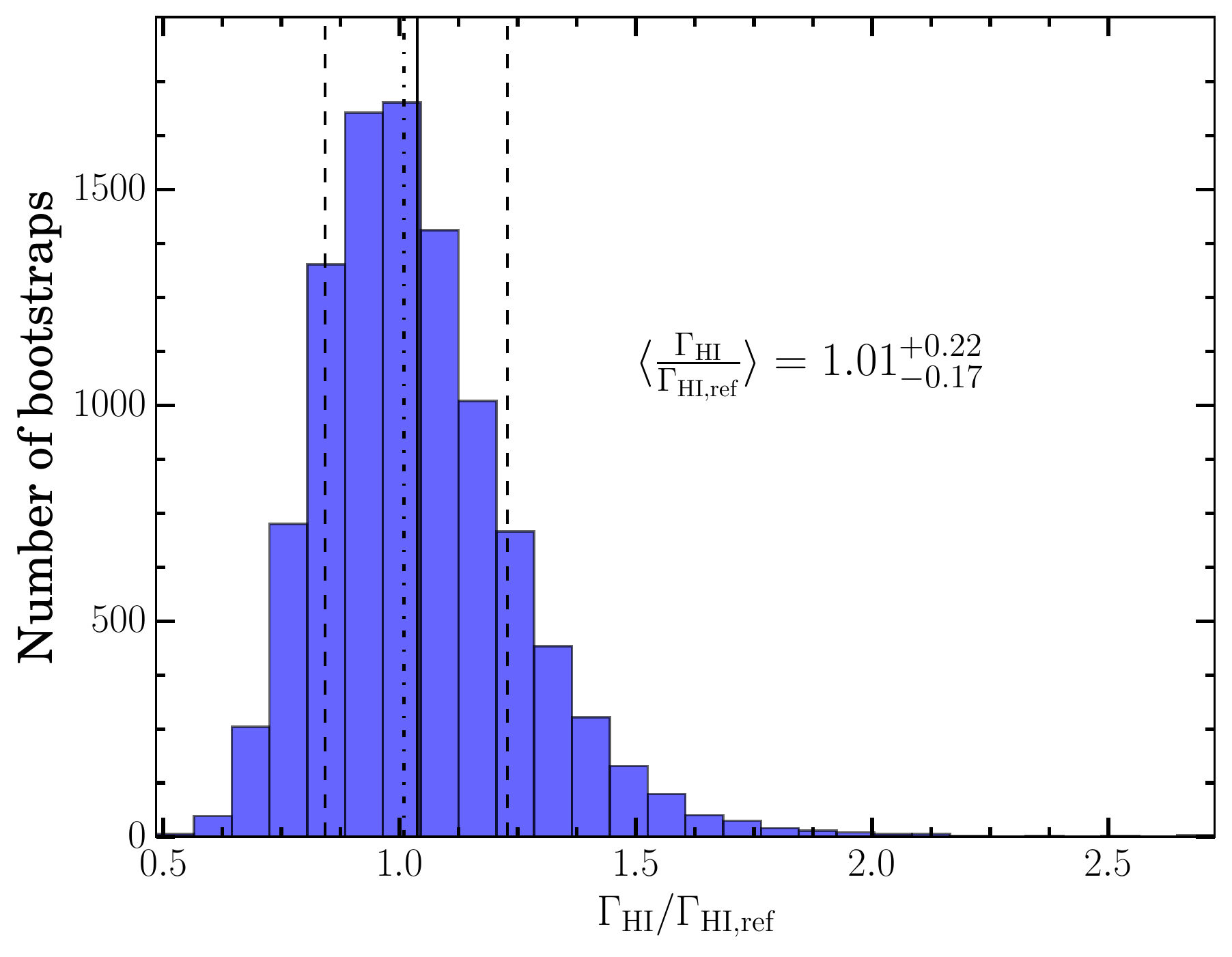}
 \caption{Histogram of the ratio of \ion{H}{i} photoionisation rates $\Gamma_\textrm{\ion{H}{i}}/\Gamma_\textrm{\ion{H}{i},ref}=3.2/\left(1.5-\alpha_\mathrm{EUV}\right)$. The mean, the median, the 16$^{\rm th}$ and the 84$^{\rm th}$ percentile are plotted as solid, dot-dashed, and dashed lines, respectively}
 \label{hist_gamma}
\end{figure}

Nevertheless, Equation~\eqref{defgammaq} provides an estimate on the uncertainty of $\Gamma_\textrm{\ion{H}{i}}$ induced by the significant uncertainty in the mean EUV spectral slope ($\alpha_\mathrm{EUV}=-1.70\pm 0.61$, Section~\ref{Spectral fit and emission lines}). For any fixed $\epsilon_{\nu,912}$ and $z$, the ratio $\Gamma_\textrm{\ion{H}{i}}/\Gamma_\textrm{\ion{H}{i},ref}=3.2/\left(1.5-\alpha_\mathrm{EUV}\right)$ quantifies the variation of the bootstrap realizations of $\Gamma_\textrm{\ion{H}{i}}$ with respect to a reference value $\Gamma_\textrm{\ion{H}{i},ref}$ that uses the slope $\alpha_\mathrm{EUV}=-1.7$ determined from the stack. As shown in Fig.~\ref{hist_gamma}, the uncertain EUV slope results in a $\sim 20$\% uncertainty in $\Gamma_\textrm{\ion{H}{i}}$. We conclude that detailed photoionisation models of the IGM need to take into account the uncertainty in the mean EUV spectral index that is dominated by the large IGM correction in our study, but dominated by sample variance in previous work (T02, S14).

Similarly, the uncertainty in $\alpha_\mathrm{EUV}$ affects estimates of the \ion{He}{ii} Lyman limit emissivity of quasars in models of \ion{He}{ii} reionization and the post-reionization UV background. Extrapolating our power-law fit to the \ion{He}{ii} Lyman limit, we obtain flux ratios $f_{\nu,1450}/f_{\nu,228}=13.8^{+18.2}_{-7.5}$ and $f_{\nu,912}/f_{\nu,228}=10.0_{-5.4}^{+14.2}$, respectively. Currently, the mean quasar SED is poorly constrained at $\lambda<500$\,\AA\ due to small sample size ($<10$ quasars) and incomplete IGM correction in the T02 sample. As the \ion{He}{ii} photoionisation rate depends on the unconstrained spectral slope at $\lambda<228$\,\AA, models of the \ion{He}{ii}-ionising background are highly uncertain. 

\rev{Lastly, we emphasize that any broken power law does not have any physical grounds, and thus it is a very poor description of the quasar SED. In principle, one should consider the full spectral information, although spectra do not usually cover energies well beyond 1 Ryd.}

\section{Summary and Conclusions}
\label{Summary and Conclusions}

We presented the first ultraviolet average spectrum at high redshift ($\langle z_{\rm em} \rangle\sim2.44$) with proper correction for the intervening \ion{Ly}{$\alpha$} forest and continuum absorption by employing the state-of the-art IGM transmission function which have been calibrated from multiple quasar absorption line observables. The sample consists of 53 quasars selected from the \hst survey for Lyman limit absorption systems (LLS) using the Wide Field Camera 3 (WFC3) presented in \citet{2011ApJS..195...16O}.

The rest-frame continuum slope of the full sample shows a break at around 912\AA\ with a FUV spectral index $\alpha_\nu=-0.61\pm0.01$ and a softening at shorter wavelengths ($\alpha_\nu=-1.70 \pm 0.61$).

Our analysis highlights the fact that slope and spectral break in the HST composite reported by T02 are incorrect due to an underestimated intergalactic absorption correction, especially at short wavelength where high-z quasars are contributing.
The FUSE composite by S04 might be less affected by this problem given the low average redshift of their sample ($z\sim0.16$). The COS composite by S12 (and S14) does not significantly differ from our spectrum within the uncertainties.

Our observed broad line ratios are in good agreement with those predicted by the photoionisation models discussed in Baskin et al. (2014), where the input quasar continuum has a EUV slope consistent with the one we observed. 

The accretion disc+X--ray corona models constructed by \citet{2012MNRAS.420.1848D} show that we need to have non zero values of the BH spin to match the data, with higher values of the BH spin as the $\mbh$ increases (assuming that the BH masses derived from \ion{C}{iv} are correct).
We do not need BH spin if we instead consider a factor of 2 lower masses (on average).
The latter is in agreement with previous results that have found BH mass values from \ion{C}{iv} systematically higher once calibrated with virial mass estimators based on the \ion{Mg}{ii} (e.g. \citealt{2008ApJ...680..169S}, but see also \citealt{2013MNRAS.434..848R}). 

Finally, we found that the lack of knowledge of the spectral slope results in an uncertainty on the quasar photoionisation rate of the order of 20\%. This effect should be taken into account by any photoionisation model.

As a final comment, we want to stress that all previous analyses were based on quasar samples selected with an unknown selection function (i.e. the best from the archive), and were characterised by small samples sizes, especially at short wavelengths (less than 10 AGN observations cover the wavelength range between $450-600$\AA~ in the S04 sample, and between 10-20 AGN in the T02 sample, and only 6 contributing at 600\AA in the S12 sample). Furthermore, errors on the composite were not published, so an assessment of the sample variance is impossible.  A single statistical correction for the IGM absorption should not be applied for a broad range of redshifts and to single spectra, since LLSs cannot all be identified or corrected for, especially not the partial LLSs.

\section*{Acknowledgements}
\rev{We would like to thank the referee for his/her helpful comments.}
We would like to thank David Schiminovich for delivering us the GALEX forced photometry prior publication.
We are grateful to Michael Shull and Jennifer E. Scott for providing us their ultraviolet composites, and Gerard Kriss for information on the T02 sample. 
EL thanks Len Cowie for clarifications on the SED shown in Fig.~\ref{sed_comp}.
We also thank the members of the ENIGMA group\footnote{http://www.mpia-hd.mpg.de/ENIGMA/} at the Max Planck Institute for Astronomy (MPIA) and Guido Risaliti for helpful discussions.
JFH acknowledges generous support from the Alexander  Humboldt foundation in the context of the Sofja Kovalevskaja Award. 
JXP acknowledges support from the National Science Foundation (NSF) grant AST-1010004 and thanks the Alexander von Humboldt foundation for a visitor fellowship to the MPIA where part of this work was performed, as well as the MPIA for hospitality during his visits.
Based on observations made with the NASA/ESA Hubble Space Telescope, obtained at the Space Telescope Science Institute, which is operated by the Association of Universities for Research in Astronomy, Inc., under NASA contract NAS 5-26555. These observations are associated with program 11594.
\rev{We have made use of the ROSAT Data Archive of the Max-Planck-Institut f\"{u}r extraterrestrische Physik (MPE) at Garching, Germany. This work is partially based on observations obtained with XMM-Newton, an ESA science mission with instruments and contributions directly funded by ESA Member States and the USA (NASA). 
The scientific results reported in this article are based in part on data obtained from the Chandra Data Archive.}

\appendix

\section{Multiwavelength properties of the WFC3 quasar sample}
\label{appendixA}
\rev{We listed the multi wavelength properties of the WFC3 quasar sample in Tables~\ref{tbl-x}, \ref{tbl-r}, and \ref{tbl-g}}.

\begin{table*}
\centering
\caption{RASS properties of the WFC3 quasar sample. \label{tbl-x}} 
\begin{tabular}{c c  l c l c}

 Name & z  & Exp.$^{\mathrm{a}}$  & $\nh$$^{\mathrm{b}}$    &  F$_{[0.5-2]\rm keV}$$^{\mathrm{c}}$ & L$_{[0.5-2]\rm keV}$$^{\mathrm{d}}$ \\ [0.5ex]
                                      &          & s               &  cm$^{-2}$ &   erg s$^{-1}$cm$^{-2}$ & erg s$^{-1}$ \\ [0.5ex]
\hline\noalign{\smallskip}
 J0755+2204 & 2.319  & 23963 & 5.73 & 4.25  $\pm$ 1.76  & 1.87   $\pm$ 0.77 \\
 J1107+6420 & 2.316  & 7728  & 1.00 & 6.74  $\pm$ 1.11  & 2.95   $\pm$ 0.48 \\
 J1119+1302 & 2.394  & 20105 & 1.96 & 9.16  $\pm$ 1.93  & 4.33   $\pm$ 0.98 \\
 J1235+6301 & 2.383  & 3311  & 1.21 & 5.56  $\pm$ 1.45  & 2.61   $\pm$ 0.68 \\
 J1253+0516 & 2.398  & 201   & 2.17 & 38.66 $\pm$ 15.21 & 18.41  $\pm$ 7.24 \\
 J1335+4542 & 2.452  & 616   & 1.88 & 10.94 $\pm$ 4.80  & 5.50   $\pm$ 2.41 \\
 J1415+3706 & 2.374  & 18476 & 0.91 & 9.47  $\pm$ 1.11  & 4.40   $\pm$ 0.52 \\
 J1454+0325 & 2.368  & 28496 & 3.62 & 24.44 $\pm$ 2.70  & 11.24  $\pm$ 0.96 \\ 
 J1625+2646 & 2.518  & 5006  & 3.37 & 14.40 $\pm$ 2.04  & 7.73   $\pm$ 1.09 \\
 J1651+4002 & 2.343  & 7649  & 1.45 & 17.21 $\pm$ 2.29  & 7.74   $\pm$ 1.03 \\
 \hline\hline
\end{tabular}

\flushleft\begin{list}{}
 \item[${\mathrm{a}}$]{ Exposure time.}
 \item[][${\mathrm{b}}$]{ Galactic column density ($\times10^{20}$) \citep{2005A&A...440..775K}.}
 \item[][${\mathrm{c}}$]{ X--ray fluxes ($\times10^{-14}$) at 0.5--2 keV calculated from the count rates in the ROSAT band using a power law spectrum with a photon index $\Gamma=2.0$ corrected for Galactic absorption \citep{2005A&A...440..775K}.}
 \item[][${\mathrm{d}}$]{ Rest frame luminosity ($\times10^{45}$) in the 0.5--2 keV band (estimated from the unabsorbed X-ray flux).}
\end{list}
\end{table*}

\begin{table*}
\centering
\caption{Radio properties of the WFC3 quasar sample. \label{tbl-r}} 
\begin{tabular}{@{}l r c c c c r}
 Name  &    F$_{\rm int}^{\mathrm{a}}$  &  Dist.$^{\mathrm{b}}$ & $f_{6{\rm cm}}^{\mathrm{c}}$  & log$f_{2500}^{\mathrm{d}}$ & $M_{2500}^{\mathrm{e}}$ & R$^{\mathrm{f}}$ \\ [0.5ex]
\hline\noalign{\smallskip}
J0806+5041   &  19.66   &    0.3  &   20.02 &  -26.45 &  -28.96 &    57.08 \\
J0854+5327   &  22.07   &    0.2  &   22.35 &  -26.04 &  -29.96 &    24.52 \\
J1119+1302   &  12.97   &    0.4  &   23.59 &  -26.58 &  -28.57 &    90.48 \\
J1335+4542   & 269.05   &    0.1  &  273.80 &  -26.62 &  -28.53 &  1150.32 \\
J1342+6015   &   1.82   &    1.3  &    1.83 &  -26.51 &  -28.76 &     5.90 \\
J1415+3706   & 409.69   &    0.3  &  412.16 &  -26.38 &  -29.05 &   995.75 \\
J1454+0324   &   4.92   &    0.2  &    4.95 &  -26.48 &  -28.79 &    15.09 \\
J1535+4836   & 107.81   &    0.1  &  111.14 &  -26.46 &  -29.04 &   321.50 \\
J1540+4138   &  18.28   &    0.3  &   18.77 &  -26.28 &  -29.45 &    36.22 \\
J1625+2646   &   9.69   &    0.1  &    9.95 &  -26.37 &  -29.08 &    23.11 \\
J1651+4002   &  43.90   &    0.5  &   43.96 &  -26.28 &  -29.28 &    83.07 \\
J2338+1504   &  45.70   &    0.5  &   43.75 &  -26.36 &  -29.15 &  101.30 \\
 \hline\hline
\end{tabular}

\begin{list}{}
 \item[${\mathrm{a}}$]{ Integrated flux density for source in mJy.}
 \item[][${\mathrm{b}}$]{ Distance from search position to source in arcsec.}
 \item[][${\mathrm{c}}$]{ Observed radio flux density at rest-frame 6 cm in mJy. }
 \item[][${\mathrm{d}}$]{ Observed optical flux density at the rest frame 2500\AA~in log erg s$^{-1}$cm$^{-2}$Hz$^{-1}$.} 
 \item[][${\mathrm{e}}$]{ Absolute magnitude at rest-frame 2500\AA~ calculated from $f_{2500}$. }
 \item[][${\mathrm{f}}$]{ Radio loudness defined as $R= f_{6{\rm cm}}/f_{2500}$.}
\end{list}
\end{table*}

\begin{table*}
\centering
\caption{GALEX forced-photometry of the WFC3 quasar sample. \label{tbl-g}} 
\begin{tabular}{@{}c c c c c c}
 Name  &  z &  NUV$^{\mathrm{a}}$  &  NUV$_{\mathrm{IVAR}}^{\mathrm{b}}$ & FUV$^{\mathrm{a}}$ & FUV$_{\mathrm{IVAR}}^{\mathrm{b}}$  \\ [0.5ex]
\hline\noalign{\smallskip}
  J0751+4245 & 2.453  &   0        &   0                       &      0        &   0          \\     
  J0755+2204 & 2.319  &   10.702   &   19.899                  &      -0.463   &   29.254     \\     
  J0806+5041 & 2.457  &   2.085    &   0.001                   &      0.052    &   0.679      \\     
  J0833+0815 & 2.581  &   0.067    &   0.649                   &      0.291    &   10.986     \\     
  J0845+0722 & 2.307  &   9.277    &   0.155                   &      -0.103   &   41.101     \\     
  J0850+5636 & 2.464  &   0.365    &   19.704                  &      0.673    &   9.939      \\     
  J0853+4456 & 2.54   &   0.248    &   16.721                  &      -0.094   &   27.238     \\     
  J0854+5327 & 2.418  &   0.734    &   15.757                  &      -0.444   &   2.087      \\     
  J0909+0415 & 2.444  &   -0.151   &   23.423                  &      0.151    &   9.292      \\     
  J0949+0522 & 2.316  &   0.351    &   18.34                   &      -0.342   &   112.288    \\     
  J1005+5705 & 2.308  &   0.25     &   0.713                   &      -0.313   &   61.293     \\     
  J1011+0312 & 2.458  &   0.252    &   1.919                   &      0.028    &   21.285     \\     
  J1053+4007 & 2.482  &   5.892    &   0.441                   &      0.178    &   10.585     \\     
  J1104+0246 & 2.532  &   2.159    &   0.032                   &      -0.011   &   2453.64    \\     
  J1107+6420 & 2.316  &   81.95    &   4.905$\times10^{-5}$    &      50.28    &   3.828$\times10^{-5}$          \\     
  J1119+1302 & 2.394  &   0.475    &   35.985                  &      0.044    &   51.255     \\     
  J1135+4607 & 2.496  &   0.355    &   4.004                   &      0.486    &   12.893     \\     
  J1143+0524 & 2.561  &   9.468    &   644.381                 &      0.008    &   5527.879   \\     
  J1215+4248 & 2.31   &   3.616    &   4.019                   &      0.015    &   3909.82    \\   
  J1220+4608 & 2.446  &   12.665   &   1.711                   &      14.278   &   0.894      \\   
  J1228+5107 & 2.45   &   23.419   &   1.31                    &      -0.002   &   33.744     \\   
  J1235+6301 & 2.384  &   106.753  &   0.107                   &      0.133    &   35.543     \\   
  J1248+5809 & 2.599  &   1.221    &   0.001                   &      -0.023   &   287.568    \\   
  J1253+0516 & 2.398  &   -0.158   &   67.674                  &      0.146    &   43.164     \\   
  J1259+6720 & 2.443  &   -0.274   &   4.891                   &      0.427    &   7.997      \\   
  J1300+0556 & 2.446  &   0.764    &   1461.025                &      0.031    &   5611.715   \\   
  J1302+0254 & 2.414  &   1.006    &   10.751                  &      0.018    &   85.795     \\   
  J1311+4531 & 2.403  &   3.621    &   1.849                   &      -0.174   &   7.709      \\   
  J1318+5312 & 2.321  &   0.372    &   29.747                  &      0.126    &   31.153     \\   
  J1323+4149 & 2.44   &   9.324    &   0.009                   &      5.766    &   0.022      \\   
  J1325+6634 & 2.511  &   0.118    &   0.121                   &      -0.089   &   499.471    \\   
  J1334+0355 & 2.583  &   1.219    &   0.031                   &      -0.248   &   112.13     \\   
  J1335+4542 & 2.452  &   1.309    &   13.318                  &      0.459    &   11.939     \\   
  J1335+4637 & 2.474  &   2.82     &   4.541                   &      0.583    &   6.58       \\   
  J1336+0157 & 2.379  &   0.378    &   7.184                   &      0.111    &   15.703     \\   
  J1342+6015 & 2.399  &   -0.158   &   35.309                  &      0.077    &   18.418     \\   
  J1354+5421 & 2.294  &   0        &   0                       &      0        &   0          \\   
  J1354+0020 & 2.504  &   0.134    &   9.851                   &      -0.055   &   43.665     \\   
  J1358+0505 & 2.455  &   0.125    &   4328.555                &      -0.013   &   5909.892   \\   
  J1400+6430 & 2.359  &   7.501    &   0.169                   &      2.517    &   0.112      \\   
  J1415+3706 & 2.374  &   3.522    &   1.408                   &      -0.067   &   497.769    \\   
  J1454+0324 & 2.368  &   2.602    &   11.792                  &      -0.168   &   126.83     \\   
  J1533+3843 & 2.529  &   0.108    &   30.193                  &      0.051    &   22.415     \\   
  J1535+4836 & 2.542  &   1.847    &   5.627                   &      0.239    &   1.424      \\   
  J1540+4138 & 2.516  &   0        &   0                       &      0        &   0          \\   
  J1610+4423 & 2.588  &   1.151    &   6.765                   &      -0.243   &   13.434     \\   
  J1625+2943 & 2.357  &   -0.054   &   12.542                  &      0.233    &   1.771      \\   
  J1625+2646 & 2.518  &   57.894   &   0.382                   &      0.267    &   15.956     \\   
  J1651+4002 & 2.343  &   0.184    &   11.125                  &      0.189    &   108.886    \\   
  J1724+5314 & 2.547  &   0.608    &   63.628                  &      -0.206   &   230.608    \\   
  J2111+0024 & 2.325  &   0        &   0                       &      0        &   0          \\   
  J2136+1029 & 2.555  &   0.276    &   0.643                   &      0.137    &   6.969      \\   
  J2338+1504 & 2.419  &   4.765    &   0.101                   &      0.435    &   0.234      \\ 
 \hline\hline
\end{tabular}

\flushleft\begin{list}{}
 \item[${\mathrm{a}}$]{ Fluxes in AB nanomaggies (monochromatic AB fluxes are defined as $F_{\rm AB} = F_{\rm AB, nano} \times10^{-71.1\times0.4}$ ).}
 \item[][${\mathrm{b}}$]{ Pipeline inverse variance (IVAR) in AB nanomaggies.}

\end{list}
\end{table*}

\section{Absolute magnitude relations}
\label{appendixB}
Below we provide some useful relations between absolute magnitudes at various wavelengths.
$\mi$ is the mean absolute i-band magnitude normalised at $z=2$ (Shen et al. 2011), $\mnorm$ and $\mryd$ are the absolute magnitudes at 1450\AA\ and 912\AA\, respectively, estimated from our WFC3 stack. 
\begin{equation}
\mi - \mnorm = -1.28
\end{equation}
\begin{equation}
\mryd - \mnorm = 0.33
\end{equation}
All these absolute magnitudes are estimated from our average WFC3 spectrum rather accurately. We finally note that, typically, the ionising luminosity is modelled as a power law, thus our estimated $\alpha_{\rm EUV}=-1.7$ can be used to evaluate the ionising luminosity as 
\begin{equation}
L_\nu = L_{912} \left(\frac{\nu}{\nu_{912}}\right)^{\alpha_{\rm EUV}},
\end{equation}
where the normalization, $L_{912}$, is well determined. 

\bibliographystyle{mn2e}
\bibliography{bibl}

\label{lastpage}
\end{document}